\documentclass[aip,amsmath,amssymb,onecolumn,notitlepage,
reprint,%
]{revtex4-2}
\usepackage{graphicx}% Include figure files
\usepackage{dcolumn}% Align table columns on decimal point
\usepackage{bm}% bold math
\usepackage[utf8]{inputenc}
\usepackage[T1]{fontenc}
\usepackage{mathptmx}
\usepackage[normalem]{ulem} 	%need for strikethrough
\usepackage[usenames,dvipsnames]{color}
\usepackage{xcolor} % for gray
\usepackage{SIunits} % required for a nice micrometer

\newif\iffinal
\finalfalse  % comment this to remove comments in output, Ignore latex errors.

\newcommand{\del}[1]{\sloppy{\textcolor{blue}{\sout{#1}}}} % delete this
\newcommand{\macom}[1]{{\marginpar{\textcolor{blue}{#1}}}} % comment on margin without text highlight

\newcommand{\intversion}[1]{} % don't show in this version

\iffinal
	\renewcommand{\del}[1]{}

	\renewcommand{\macom}[1]{}
	
\else
\fi

\makeatletter
\let\@fnsymbol\@fnsymbol@latex
\@booleanfalse\altaffilletter@sw
\makeatother
\begin{document}

%////////////////
\title{Enhancing the Coherence of Superconducting Quantum Bits with Electric Fields}

\author{J\"urgen Lisenfeld}\thanks{Corresponding author: juergen.lisenfeld@kit.edu}
\affiliation{Physikalisches Institut, Karlsruhe Institute of Technology, 76131 Karlsruhe, Germany}
\author{Alexander Bilmes}
\affiliation{Physikalisches Institut, Karlsruhe Institute of Technology, 76131 Karlsruhe, Germany}
%\affiliation{Google, Santa Barbara, USA}
%\author{Alexander Konstantin Neumann}
%\affiliation{Physikalisches Institut, Karlsruhe Institute of Technology, 76131 Karlsruhe, Germany}
\author{Alexey V. Ustinov}
\affiliation{Physikalisches Institut, Karlsruhe Institute of Technology, 76131 Karlsruhe, Germany}
%\affiliation{National University of Science and Technology MISIS, Moscow 119049, Russia}
%\affiliation{Russian Quantum Center, Skolkovo, Moscow 143025, Russia}

\date{\today}

\begin{abstract}
	\centering\begin{minipage}{\linewidth}
		\textbf{
  			In the endeavour to make quantum computers a reality, 
	      	integrated superconducting circuits have become a promising architecture.
			A major challenge of this approach is decoherence originating from spurious atomic tunneling defects at the interfaces of qubit electrodes, which may resonantly absorb energy from the qubit's oscillating electric field and reduce the qubit's energy relaxation time $T_1$.
			Here, we show that qubit coherence can be improved by tuning dominating defects away from the qubit resonance using an applied DC-electric field.	
			We demonstrate a method that optimizes the applied field bias and enhances the 30-minute averaged qubit $T_1$ time by 23\%.
			We also discuss how local gate electrodes can be implemented in superconducting quantum processors to enable simultaneous in-situ coherence optimization of individual qubits.		
	}
	\end{minipage}
\end{abstract}

\maketitle 
\setlength{\parskip}{-0.25cm}

\section*{Introduction}

\begin{figure*}[htb!]
	\begin{center}
		\includegraphics[width=.99\textwidth]{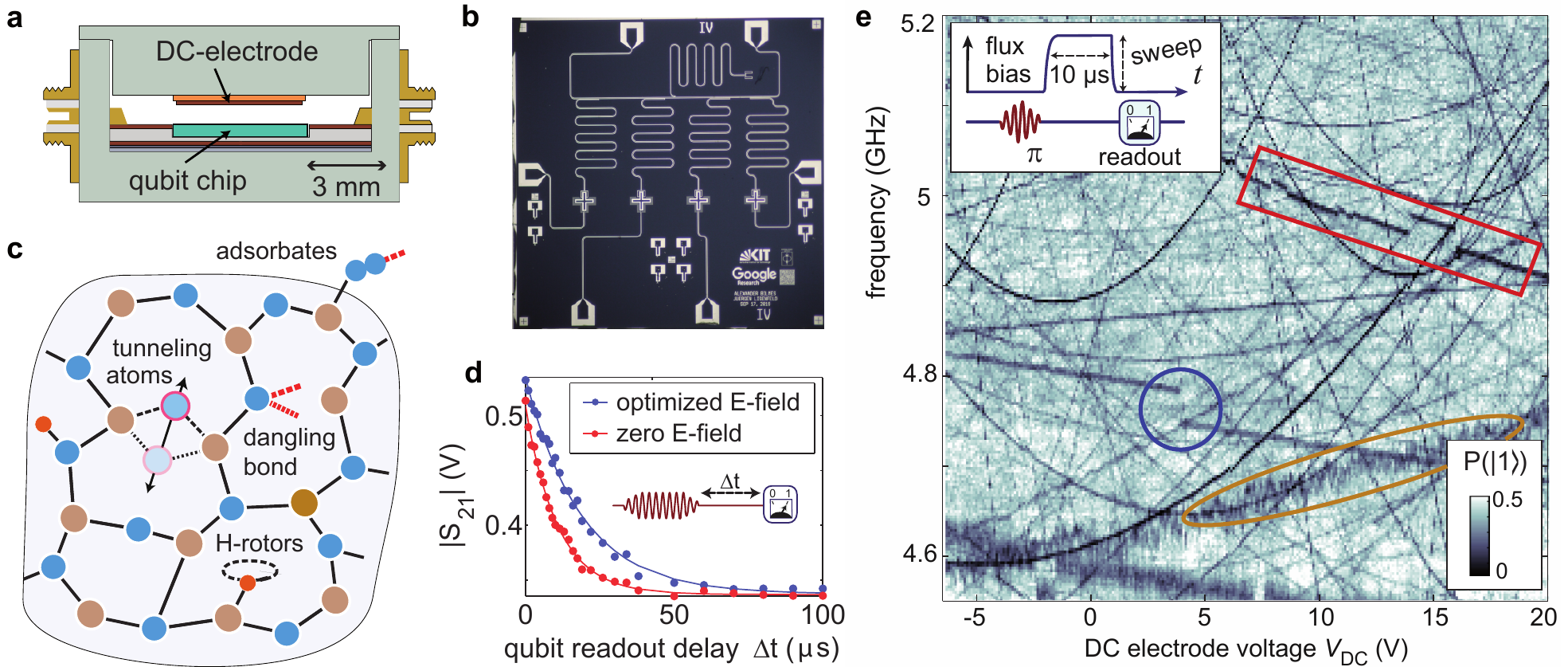} 
	\end{center}
	\caption{\textbf{Tuning defects by an electric field.}
		\textbf{a}~Cross-section through the sample housing. The electrode to generate the E-field consists of a Copper-foil/Kapton foil stack glued to the lid of the sample holder above the qubit chip, and voltage-biased against ground. \textbf{b} Photograph of the XMon qubit samples used in this work. 
		\textbf{c} Illustration of defects which appear in the amorphous oxides of qubit electrodes.
		\textbf{d} Exemplary measurements of the decaying qubit population after a long exciting microwave pulse (see inset) to determine the $T_1$ time. Red (blue) points were acquired at zero (the optimized) applied E-field.
		\textbf{e} Resonances of individual TLS (dark traces), observed as accelerated decay of the qubit's excited state population (colour scale) using the swap-spectroscopy protocol shown in the inset. The circle marks coupling of a TLS to a metastable fluctuator which may cause hysteresis in E-field sweeps. Rectangle and ellipse indicate the fluctuating resonance frequencies of TLS coupled to slowly and quickly fluctuating thermal TLSs, respectively.		
	}
	\label{fig:1}
\end{figure*}

Superconducting integrated circuits have evolved into a powerful architecture for creating artificial quantum systems. In state-of-the-art experiments, tens of qubits are coherently operated as quantum simulators and universal processors\cite{arute2019quantum,IBMqv64,Rigetti19,wu2021} while access to prototype devices is being offered via the cloud to accelerate the development of practical quantum algorithms\cite{quantumclouds}. On the way forward, mitigating decoherence is one of the central challenges, because it hinders further up-scaling and implementation of quantum error correction\cite{murray2021,kjaergaard2019}.\\

Today's processors typically employ transmon qubits that are based on discrete energy levels in non-linear LC-resonators formed by a capacitively shunted Josephson junction\cite{KochTransmon}.
%, which consist of a capacitor shunted by a Josephson junction to realize a nonlinear LC resonator whose discrete energy levels can be selectively addressed by resonant microwave pulses.
A large part of decoherence in such qubits is due to dielectric loss in the native surface oxides of the capacitor electrodes\cite{Wang2015,Lisenfeld19}. This loss shows a remarkably structured frequency dependence~\cite{Martinis:PRL:2005,Barends13} which originates in the individual resonances of spurious atomic tunneling defects\cite{Muller:2017}.
These defects form a sparse bath of parasitic two-level quantum systems, so-called TLS, which have been evoked long ago to explain the anomalous low-temperature properties of amorphous materials~\cite{Phillips87,Anderson:PhilMag:1972}.%,  an effort that is still ongoing.
When a TLS has an electric dipole moment, it may resonantly absorb energy from the oscillating electric field of the qubit mode, and efficiently dissipate it into the phonon-~\cite{Jaeckle72} or BCS quasiparticle bath~\cite{Bilmes17}.
Moreover, TLS resonance frequencies may fluctuate in time due to interactions with thermally activated, randomly switching low-energy TLS\cite{Black:PRB:1977, klimov2018,Schloer2019,burnett2019,carroll2021}. 
This mechanism efficiently transforms thermal noise into the qubit's environmental spectrum, and causes fluctuations of the qubit's resonance frequency and energy relaxation rate $T_1$~\cite{Faoro:PRL:2012,Mueller:2014,FaoroResonators}. For quantum processors, this implies fluctuations of their quantum volume (i.e. computational power)\cite{pelofske2022}.
\\
% note that qubits may shed light on long-standing issues regarding the anomalies of glasses
% properties and material research using resonators and qubits.
% quantum tomography of TLS coherence, environmental spectroscopy, NMR control, QP interactions, mutual TLS interactions

Recently, we have shown that the resonance frequencies of TLS located on thin-film electrodes and the substrate of a qubit circuit can be tuned by an applied DC-electric field~\cite{Lisenfeld19,Bilmes20}. Accordingly, it becomes possible to tune defects that dominate qubit energy relaxation away from the qubit resonance, and this results in longer relaxation times $T_1$.
Here, we demonstrate this concept using a simple routine which maximizes the $T_1$ time of a qubit by searching for an optimal electric field bias. The method was tested at various qubit resonance frequencies and increased the 30-minute averaged qubit $T_1$ time by 23\%.\\

%This is useful to resolve the positions of individual TLS along the circuit electrodes, which can guide efforts to avoid defect formation by improved fabrication methods.\\
%Here, we show that electric-field tuning of TLS allows one to enhance the energy relaxation rate of a qubit. We demonstrate a simple routine that optimizes the qubit's $T_1$ time by tuning defects reliably out of qubit resonance. 

\section*{Electric field tuning of TLS}
For our experiments, we fabricated a transmon qubit sample in the so-called 'X-Mon' design following Barends et al.~\cite{Barends13} as shown in Fig.~\ref{fig:1}\,\textbf{b}. The flux-tunable qubit uses a submicron-sized Al/AlOx/Al tunnel junction made by shadow evaporation as described in detail in Ref.~\cite{Bilmes22strayjj}.
The electric field for TLS tuning is generated by a DC-electrode installed on the lid of the sample housing $\approx 0.9~\mathrm{mm}$ above the qubit chip's surface as illustrated in Fig.~\ref{fig:1}\textbf{a}. The electrode is made from a copper foil that is insulated by Kapton foil from the housing. To improve E-field homogeneity in vicinity of the qubits, the electrode has a comparable size than the qubit chip. More details on this setup are described in Ref.~\cite{Lisenfeld19}.\\

The response of TLS to the applied electric field is observed by measuring the qubit energy relaxation time $T_1$ as a function of qubit frequency, which shows Lorentzian minima whenever sufficiently strongly interacting TLS are tuned into resonance. A detailed view on the rich TLS spectrum as shown in Fig.~\ref{fig:1}\,\textbf{e} is obtained using swap-spectroscopy~\cite{Lisenfeld2015}. With this protocol, TLS are detected by the resonant reduction of the qubit's excited state population after it was tuned for a fixed time interval to various probing frequencies. In the studied sample, only a single TLS was observed that did not couple to the applied E-field, indicating that it was likely residing in a tunnel barrier of the submicron-sized qubit junctions where no DC-electric field exists~\cite{KochTransmon}. This confirms that only a few resonant TLS are typically found in small area Josephson junctions\cite{kim2008, murray2021,Bilmes22strayjj}, and dielectric loss is dominated by defects on the interfaces of the qubit electrodes\cite{Lisenfeld19, Bilmes20, Wang2015}. This is true as long as qubits are fabricated with methods~\cite{dunsworth2017,Bylander2020,Bilmes21} that avoid the formation of large-area stray Josephson junctions which are known to contribute many additional defects~\cite{Lisenfeld19,Bilmes22strayjj}.\\

In Fig.~\ref{fig:1}\,\textbf{e}, some TLS are observed whose resonance frequencies show strong fluctuations or telegraphic switching due to their interaction with low-energy TLS that are thermally activated.
We note that TLS may also interact with classical bistable charge fluctuators that have a very small switching rate between their states. Since these fluctuators may also be tuned by the applied electric field, hysteresis effects may appear in electric field sweeps since the state of a fluctuator, and hereby the resonance frequency of a high-energy TLS, may depend on the history of applied E-fields~\cite{Meissner18}. An example of such an interacting TLS-fluctuator system is marked by the blue circle in Fig.~\ref{fig:1}\,\textbf{e}, where the resonance frequency of a TLS abruptly changed.\\

\section*{Method for optimizing the qubit $T_1$ time}
As it is evident from Fig.~\ref{fig:1}\,\textbf{e}, at each qubit operation frequency there is a preferable electric field bias where most of the dominating TLS are tuned out of qubit resonance and the $T_1$ time is maximized.
In the following, we describe a simple routine by which an optimal E-field bias can be automatically determined.\\

\begin{figure}[htb]
	%	\begin{center}
	\includegraphics[width=\columnwidth]{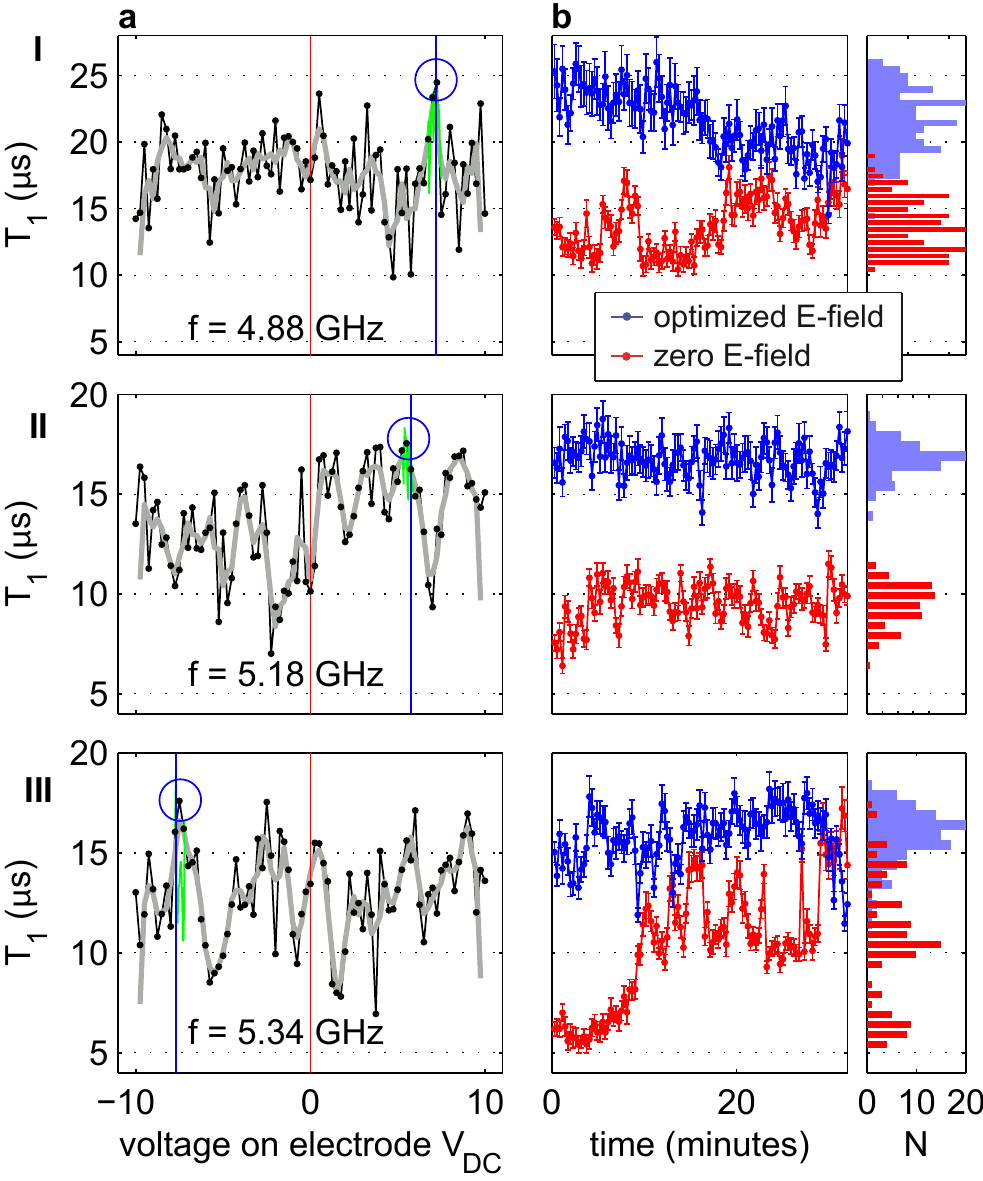} 	
	%	\end{center}
	\caption{\textbf{Benchmarking the optimization algorithm}.
		Rows \textbf{I}\ -\ \textbf{III} were taken at the indicated qubit operation frequencies. %. (cf. Fig.~\ref{fig:3}\textbf{a}).	
		\textbf{a}~Qubit $T_1$-time vs. applied electric field.
		Black data points are obtained from fits to exponential decay curves as shown in Fig.~\ref{fig:1}\,\textbf{d}. Dips in $T_1$ indicate the resonance with strongly coupled TLS. The gray curve is a 3-point nearest neighbour average, whose maximum (blue circle) is close to the determined optimum bias field. The green line indicates $T_1$ obtained in the second pass, sweeping the field in finer steps in a small range around the E-field of previously determined $T_1$ maximum. 		
		\textbf{b} Comparison of qubit $T_1$ times between an optimized E-field bias (blue data) vs. zero applied E-field (red data), measured as a function of time during 30 minutes.
}
	\label{fig:2}
\end{figure}

First, the qubit $T_1$-time is measured for a range of applied electric fields. Hereby, the $T_1$-time is obtained from exponential fits to the decaying qubit population probability after it was excited by a microwave pulse, measured using the common protocol shown in the inset of Fig.~\ref{fig:1}\,\textbf{d}.
Figure~\ref{fig:2}\textbf{a} shows the resulting electric field dependence of $T_1$ (black data points), measured at various qubit resonance frequencies (rows I to III). 
% such data (blue curve), where resonances with TLS result in local minima in the qubit's $T_1$-time.\\
These data are then smoothed by a nearest-neighbour average (gray curve) to average out individual dips and peaks in order to amplify broader maxima that promise a more stable improvement.\\

Next, the E-field is set to the value where the maximum $T_1$-time occurred (blue circle in Fig.~\ref{fig:2}\,\textbf{a}). Hereby, it is recommended to approach the detected optimal E-field from the same value where the previous E-field sweep was started. This helps to avoid the mentioned possible hysteresis effects in the TLS resonance frequencies that may occur when they are coupled to meta-stable field-tunable TLS whose state depends on the history of applied E-fields.
% A further investigation of hysteresis effects in E-field tuning experiments is in progress and will be discussed in the final manuscript. \red{Details on second pass of optimization will be added.}
Finally, a second pass is performed, sweeping the E-field in finer steps around its previously determined optimum value until the obtained $T_1$ time is close to the maximum value that was observed in the previous sweep. This ensures that hysteresis effects are better compensated and the finer step helps to avoid sharp dips that were not resolved in the first pass. Data obtained in the second pass are plotted in green in Fig.~\ref{fig:2}\,\textbf{a}).\\

\section*{Benchmarking the method}
To test the efficiency of the optimization routine, first the qubit $T_1$ is repeatedly observed during 30 minutes at zero applied electric field as a reference (red data in Fig.~\ref{fig:2}\,\textbf{b}). Afterwards, the optimization routine searches for the electric field which maximizes the qubit's coherence time by taking data as shown in Fig.~\ref{fig:2}\,\textbf{a}. The result is then checked by monitoring the $T_1$-time at the found optimal E-field during another 30 minutes (blue data in Fig.~\ref{fig:2}\,\textbf{b}).
Evidently, during most of this time, acquired $T_1$ times after optimization are higher than the reference values that were obtained at zero applied electric field.\\

\begin{figure}[htb!]
	%	\begin{center}
	%	\includegraphics[width=\columnwidth]{./figs/ImprovementLargeFont2.png}
		\includegraphics[width=\columnwidth]{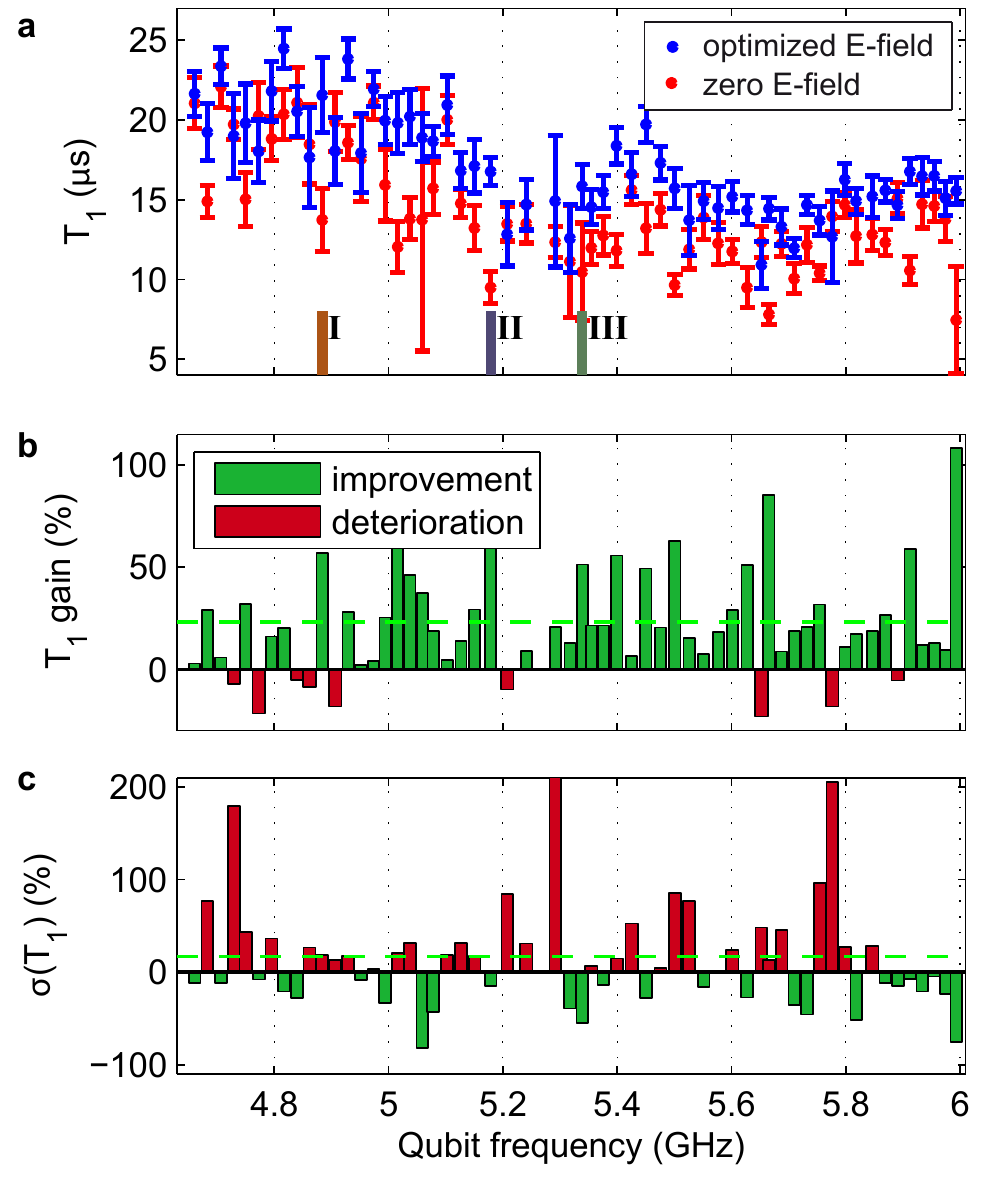}
	%	\end{center}
	\caption{\textbf{Results.} \textbf{a)} Qubit $T_1$ times after E-field optimization (blue data) and at zero applied E-field (red data), tested at various qubit frequencies and averaged over 30 minutes.
	The error bars indicate the spread (standard deviation) of $T_1$ over 30 minutes.
	\textbf{b)} Relative improvement of qubit $T_1$ after optimization. In a few cases, the routine results in a smaller $T_1$ time (red bars).  Best and average $T_1$-time improvements are 108\% and 23.2\%, respectively.
	\textbf{c)} Fluctuations of $T_1$-times (standard deviation over 30 minutes). On average, the fluctuations were 17\% higher for the optimized E-field.
}
	\label{fig:3}
	\end{figure}

To measure the average improvement of the optimization routine, the benchmarking protocol was repeated at various (in total 59) qubit resonance frequencies, see the supplementary material for the full data set.
Figures~\ref{fig:3}\textbf{a} and \textbf{b} summarize the absolute and relative improvement of the qubit $T_1$-time at all investigated qubit resonance frequencies. In most cases ($85\%$), the routine improved the 30-minute average qubit $T_1$-time. The improvement was larger than 10\% $T_1$ in 67\% of cases, and enhanced $T_1$ by more than 20\% in 46\% of all tries.\\
 
The few cases where the averaged $T_1$-time was smaller after optimization were caused by TLS resonance frequency fluctuations occurring during the 30-minute averaging interval. In quantum processors, such deterioration can be detected on the basis of qubit error rates and trigger a renewed E-field optimization.\\ 
Averaged over all tested qubit resonance frequencies and a 30-minute time interval past optimization, the $T_1$ time improvement was $\approx23\%$. We expect that similar improvements are possible also in state-of-the-art transmon qubits, as all of them show time-dependent and sample specific $T_1$ time variations which indicate their limitation by randomly occuring TLS~\cite{Mueller:2014,Muller:2017,klimov2018,Schloer2019,burnett2019}.\\

As a consequence of the defects' resonance frequency fluctuations, the enhancement (gain) of the $T_1$ time tends to diminish with time that has passed after the E-field optimization. A further analysis (see supplementary material I and II) indicates that the average $T_1$ gain drops from an initial value of about 30\% immediately after optimization to slightly above 20\% after 30 minutes past optimization.\\

To check how much the optimization routine affects the temporal fluctuation strength of the qubit's $T_1$ time, the standard deviation of observed $T_1$ times during the 30 minute intervals before and after optimization were compared. The result is shown in Figure~\ref{fig:3}\textbf{c}. In slightly more than half cases (59\%), the $T_1$ time fluctuations increased after optimization. This might be mitigated by enhancing the optimization algorithm such that it prefers broader $T_1$-time peaks which are less sensitive to TLS frequency fluctuations, and by including the $T_1$ fluctuation strength at detected peaks as a criterion.
%, showing a frequency-averaged increase of 17\%. 
\\

\section*{Proposed integration with quantum processors}
When each qubit in a processor is coupled to a dedicated local gate electrode, the optimization routine can be applied simultaneously on all qubits. This tuneup-process is facilitated when no cross-talk of a gate electrode to neighboring qubits occurs. Moreover, the generated electric field should be sufficiently strong all along the edges of the qubit island and the opposing ground plane (where surface defects are most strongly coupled to the qubit~\cite{Lisenfeld19}), so that all relevant TLS can be tuned by $\delta \varepsilon  \gtrsim 100$~MHz to decouple them from the qubit. Assuming a relatively small coupling TLS dipole moment component of $p=0.1\,\mathrm{e}$\AA~\cite{Lisenfeld19, hung2021, Martinis:PRL:2005}, this corresponds to required field strengths $E = \delta \varepsilon / p \approx 40\,$kV/m. Given a typical distance between the DC-electrode and the qubit electrodes of below 1~mm, such E-fields are unproblematically obtained with a bias voltage of a few Volt on the DC-electrode.\\

\begin{figure}[htb]
	\begin{center}
		\includegraphics[width=\linewidth]{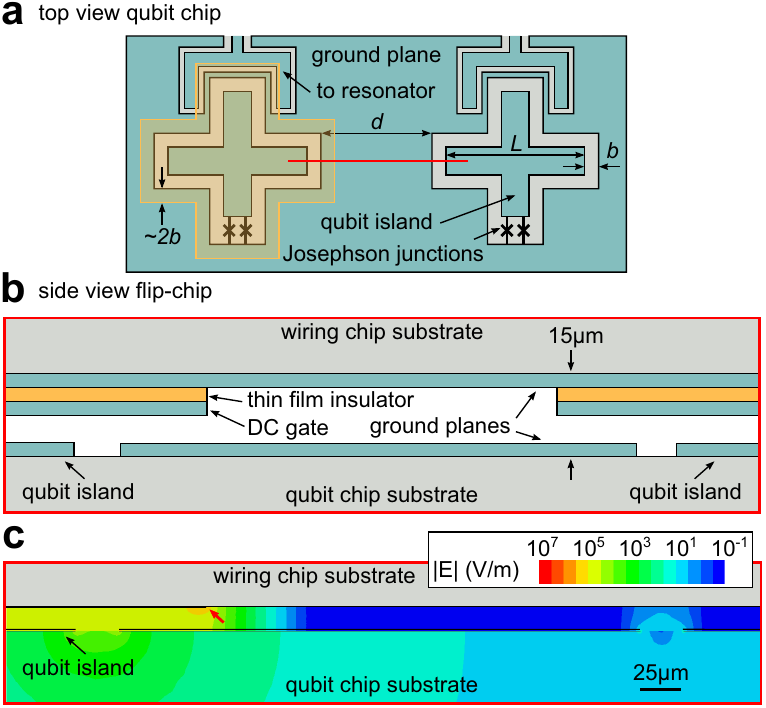}
	\end{center}
	\caption{
		\textbf{a} Top view sketch of two Xmon-qubits.
		 %The arm length of the qubit island is typically $300\,\mathrm{\mu m}$, and the gap to the surrounding ground plane typically $25\,\mathrm{\mu m}$.
		 The orange region indicates the gate electrode above the left qubit.
		\textbf{b} Cross-section of the flip-chip stack, sketched along the red line shown in \textbf{a}. The DC gate electrodes are separated by thin-film insulators from the wiring chip's ground plane.
		\textbf{c} Simulated electric field strength when $1\,\mathrm{V}$ is applied to the left gate electrode. The field decays horizontally on a scale of about the distance between the qubit and the wiring chips (here, $15\,\mathrm{\mu m})$, resulting in small cross-talk below $10^{-4}$.		
%\jl{Check E-field strength, plate cap formula gives E=1V/15um=66kV/m..}
	}
	\label{fig:gate}
\end{figure}

Figure~\ref{fig:gate} shows a possible implementation of a gate electrode array, which is located on a separate wiring chip that is bumb-bonded to the chip carrying the qubits in a flip-chip configuration~\cite{Foxen2017,kosen2022}. In Fig.~\ref{fig:gate}\,\textbf{a}, a top view of two Xmon-type~\cite{Barends13} qubits is shown, where the gate electrode above the left qubit is indicated in orange. The electrode extends slightly over the edges of the qubit island's opposing ground plane to ensure the tunability of TLS in this region.\\

The cross section of the chip stack is sketched in Fig.~\ref{fig:gate}\,\textbf{b}, showing that the gate electrodes are separated from the ground plane of the wiring chip by a thin film insulator.\\
The simulated electric field strength in this region is drawn to-scale in Fig.~\ref{fig:gate}\,\textbf{c}, for the case when the left electrode is biased at $1\,\mathrm{V}$ while all other metallic parts (including the qubit island ~\cite{Lisenfeld19}) are kept at zero potential. 
As expected, the induced field strength decays on a length scale of roughly the distance between the two chips, given that qubits are surrounded by a ground plane and also the wiring chip has a ground plane. For a qubit-to-qubit separation of $d>100\,\mathrm{\mu m}$ as used in the presented simulation, we accordingly find the cross-talk to be below $10^{-4}$.\\

Alternatively, the local electrodes could also be placed on the backside of the qubit chip. In this case, the substrate thickness will determine the horizontal field screening length, and stronger cross-talk can be expected. However, FEM simulations of the induced E-fields in a given processor layout should allow one to sufficiently compensate for this cross-talk.\\

The capacitive coupling of the qubit to the gate electrode introduces extra decoherence channels: dielectric loss occurs in the insulation separating the electrode from ground, and by radiative loss, the qubit dissipates energy into the electrode wiring. These losses depend strongly on the dimensions of the electrode.
We find (see supplementary material) a qubit $T_1$ limitation of 5 ms for the setup used in this work, and estimate similar values for the proposed integration into flip-chip quantum processors.

\section*{Conclusion}
We present an experimental setup and an automatic routine that extends the energy relaxation time $T_1$ of superconducting transmon qubits. The idea is to expose the qubit electrodes to a DC-electric field at which the most detrimental TLS-defects are tuned out of qubit resonance. Averaging over qubit working frequencies and a 30-minute time interval (that was limited by time constraints), the $T_1$-time was improved by 23\% compared to zero applied electric field.\\

%However, recent experiments~\cite{carroll2021} suggests that the frequency shifts of TLS remain smaller than 10 MHz during 10 days of measurement, which is also corroborated by the notion that each TLS can only interact with a limited number of thermally active defects in its vicinity\cite{Black:PRB:1977,Faoro:PRB:2015,Lisenfeld2016}. This suggests that our optimization routine can be further improved when it is repeatedly executed at regular intervals, in order to identify an electric field bias that protects the qubit on long timescales from interactions with strongly coupled TLS.\\ %\new{In quantum processors, the steady stream of qubit error rates available from quantum-error correction algorithms can be fed back to linear or machine learning E-field tuning algorithms to ensure continuous optimization.}

In our experiments, the optimization routine took less than 10 minutes (to acquire about 60 values of qubit $T_1$ at several E-fields). However, the data shown in Fig.~\ref{fig:2}\,\textbf{a} suggests that the range of applied E-fields may be reduced, which together with further optimizations such as less averaging in individual $T_1$-time measurements, may reduce the optimization time to about one minute.\\

Analysis of the raw data such as shown in Fig.~\ref{fig:2} and in the supplementary material suggests that more stable improvements might be achieved by improving the algorithm, e.g. by including the width of a peak in $T_1$ vs. E-field as a criterion next to the height of the peak.
Moreover, we expect that deterioration of the 30-minute average qubit $T_1$ time by the optimization routine, as it occurred in a few ($\approx~15\%$) cases in these tests, can be avoided by averaging over several E-field sweeps to better account for TLS showing strong resonance frequency fluctuations. Also, one may devise a linear or machine-learning feedback mechanism that regularly readjusts the E-field bias on the basis of the steady stream of qubit error rates obtained during quantum algorithms to ensure continuous coherence enhancement.\\

The ability to tune TLS out of resonance with a qubit is especially beneficial for processors implementing fixed-frequency qubits, which can be tuned only in a limited range by exploiting the AC-stark shift\cite{CarrollACstark}. This may still allow one to improve qubit coherence by evading strongly coupled TLS as it was recently demonstrated by Zhao et al.\cite{zhao2022}. However, even when tunable qubits are used, it is still necessary to mutually balance their individual resonance frequencies to avoid crosstalk and to maximize gate fidelities, and this will be greatly simplified if qubit coherence can be optimized at all frequencies by having independent control of the TLS bath. Also, to improve two-qubit gates that require qubit frequency excursions, one could adjust our optimization procedure to minimize the number of TLS that have resonances in the traversed frequency interval.\\

Our simulations indicated that it is straight-forward to equip each qubit in a processor with local gate electrodes, which will allow one to simultaneously improve $T_1$ of all qubits. We thus see good opportunities for this technique to become a standard in superconducting quantum processors. 
%\new{The method can also be used in other qubit implementations where qubits are sensitive to individual sparsely distributed defects that can be tuned independently from the qubit.}\jl{I now tend to skip this last paragraph - I could not think of examples other than supeconducting qubits where single defects are seen that can be tuned independently from the qubit.}
\\
 
\section*{Methods}
The qubit sample is a stray-junction free transmon qubit that was fabricated by A. Bilmes as described in detail in Ref.~\cite{Bilmes22strayjj}. For details about the experimental setup, the implementation of the DC-electrode for defect tuning, and simulations of the electric field, we refer to Ref.~\cite{Lisenfeld19}.

\section*{Data Availability}
The data that support the findings of this study are available from the corresponding author upon reasonable request.

\section*{Acknowledgements}
This work was funded by Google LLC which is gratefully acknowledged. We thank for funding from the Baden-W\"urttemberg-Stiftung (Project QuMaS), and for funding from the Bundesministerium für Forschung und Bildung in the frame of the projects QSolid and GeQCoS. We acknowledge support by the KIT-Publication Fund of the Karlsruhe Institute of Technology. 

\section*{Author contribution}
JL devised and implemented idea and method, performed the experiments, and analyzed the data. AB built the qubit setup for E-field tuning, designed and fabricated the investigated qubits, and performed FEM simulations. The manuscript was written by JL with assistance from AB.

\section*{Competing interests}
The Authors declare no competing financial or non-financial interests.

%\bibliography{Biblio}

\renewcommand*{\thesection}{S \arabic{section}}
\renewcommand{\thefigure}{S\arabic{figure}}
\setcounter{figure}{0} 
\section*{Supplementary Material}
\section{Estimation of energy losses from the global DC gate}
\noindent \textbf{Radiative loss}\\
We estimate the limitation of the qubit energy relaxation time by radiative loss into the wiring channel of the DC electric gate. Here, we discuss the case of a large ("global") gate electrode above the chip, as used in our experiment and illustrated in Supplementary Figure~\ref{fig:S20} \textbf{a}. The effective DC gate wiring diagram is shown in Supplementary Figure ~\ref{fig:S20} \textbf{c} where $C_\text{c}$ is the qubit's coupling capacitance to the DC gate, and $C_\text{f}$ the large filter capacitance of the DC gate to ground ($C_\text{c}\ll C_\text{f}$), as indicated in the panel \textbf{a}. The qubit circuit is given by the Josephson junction (defining the qubit Josephson energy $E_\text{J}$) which is connected in parallel to the qubit total shunt capacitance $C_\text{tot}=C_\text{q}+C_\text{c}\parallel C_\text{f}\sim 97\,\mathrm{fF}$ (defining the qubit charging energy $E_\text{C}=e^2/2C_\text{tot}$). Here, $C_\text{q}$ is the sum of the qubit's island capacitance to the chip's ground plane and the Josephson junction's self capacitance ($\sim 6\,\mathrm{fF}$). The copper DC wire (length $l\sim1\,\mathrm{m}$, radius $r\sim 50\,\mathrm{\mu m}$) is used to control the voltage, and it has an impedance of $Z=\sqrt{\mu_0 f_{01}/(\pi\sigma_\text{Cu})}\cdot l/(2r)\approx{400\,\mathrm{\Omega}}$~\cite{dcWireImpedance}, where $\sigma_\text{Cu}=150\cdot 10^4\,\mathrm{(\Omega m)^{-1}}$ is the specific conductivity of copper at a temperature of $2\,\mathrm{K}$ (we assume that the Cu conductivity does not change largely at lower temperatures), and $f_{01}=\sqrt{8E_\text{J}E_\text{C}}-E_\text{C}$ is the qubit resonance frequency. If the gate was connected to a standard impedance-matched coaxial RF cable, $Z$ would be $50\,\mathrm{\Omega}$.\\

\begin{figure*}[htbp]
	\begin{center}
		\includegraphics[width=.75\textwidth]{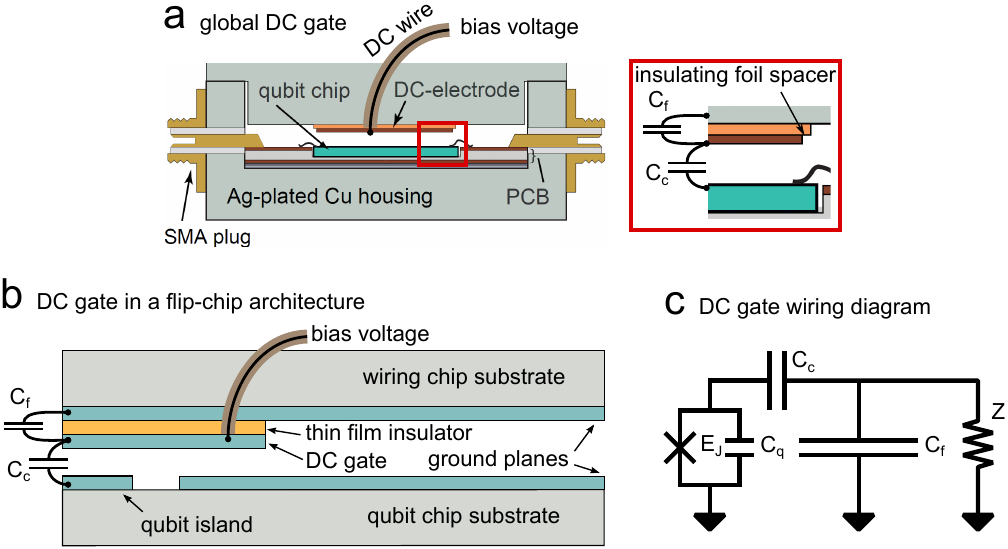} 
	\end{center}
	\caption{\textbf{The qubit and the DC gate.}
		\textbf{a} Sketch of our experimental setup - the sample box and the global DC gate electrode placed above the qubit chip.
		\textbf{b} Sketch of a local DC gate in a flip-chip architecture, as proposed in the main text.
		\textbf{c} Effective electric diagram of the qubit circuit and the DC gate. The qubit-gate coupling capacitance $C_\text{c}$ and the filter capacitance of the gate to ground $C_\text{f}$ are defined in the inset of \textbf{a} and in \textbf{b}.
	}
	\label{fig:S20}
\end{figure*}

The radiative loss of a transmon qubit which is capacitively coupled to an RF port has been calculated e.g. in the thesis by D. Sank~\cite{Sank}, see Equation (D.9):
\begin{align}
	Q_\text{l} = & \left(\frac{C_\text{tot}}{C_\text{c}}\right)^2\frac{Z_\text{q}}{\Re(Z_\text{eff})},
	\label{eq:Ql}
\end{align}
where $Q_\text{l}$ is the loaded quality factor of the qubit coupled to the RF port, and $Z_\text{q}=\sqrt{L_\text{J}/C_\text{tot}}$ is the characteristic qubit impedance. The Josephson inductance is given by $L_\text{J}=(\hbar/2e)^2/(E_\text{J}\cos{\phi})\approx(\hbar/2e)^2/E_\text{J}$, where the phase drop across the qubit Josephson junction in the Transmon qubit is $\phi\approx0$. Finally, $\Re(Z_\text {eff})$ is the real part of the RF port impedance which contributes to radiative losses of the qubit into the bandwidth of the RF port. In our case, we regard our DC gate as an RF port with an effective impedance $Z_\text{eff}=Z_\text{f}Z/(Z_\text{f}+Z)$ which results from the parallel connection of the wire impedance $Z$ and the impedance $Z_\text{f}=1/i\omega C_\text{f}$ from the filter capacitance. The radiation-limited energy relaxation time $T_{1,\text{rad}}$ is given by:
\begin{align}
	T_{1,\text{rad}} = & \frac{Q_\text{l}}{\omega_{01}},
\end{align}
where $\omega_{01}=2\pi f_{01}$ is the qubit resonance frequency.\\

\noindent \textbf{Dielectric loss}\\
Here, we account for dielectric losses in the filter capacitance $C_\text{f}$ which is indicated in Supplementary Figure~\ref{fig:S20} \textbf{a} and \textbf{c}. In our experiment we used a global gate whose dielectric is a Kapton foil (loss tangent $\tan(\delta)\sim2\cdot10^{-2}$, and $\varepsilon_\text{r}=3.5$~\cite{kapton}). The participation ratio $P_\text{f}$ of the filter capacitor is calculated as the qubit energy stored in $C_\text{f}$ divided by the total qubit energy, i.e.
\begin{align}
	P_\text{f} = & \frac{\frac{1}{2}C_\text{f}(V_\text{rms}\frac{C_\text{c}}{C_\text{c}+C_\text{f}})^2}{\frac{1}{2} \left(C_\text{q}+\frac{C_\text{c}C_\text{f}}{C_\text{c}+C_\text{f}}\right)V_\text{rms}^2}\\
	\approx & \frac{C_\text{f}(\frac{C_\text{c}}{C_\text{f}})^2}{\left(C_\text{q}+C_\text{c}\right)}\\
	= & \frac{C_\text{c}^2}{C_\text{f}\left(C_\text{q}+C_\text{c}\right)},
\end{align}
where $V_\text{rms}$ is the root mean square of the oscillating voltage across the qubit Josephson junction, or in other words the qubit's vacuum voltage fluctuations. The small portion of $V_\text{rms}$ that drops across the filter capacitor is $V_\text{rms}C_\text{c}/(C_\text{c}+C_\text{f})$. The simplified expression for $P_\text{f}$ results from $C_\text{c}\ll C_\text{f}$. The added energy loss from the filter dielectric is then
\begin{align}
	\frac{1}{T_{1,\text{diel}}} = & P_\text{f}\tan(\delta)\omega_{01},
	\label{eq:t1diel}
\end{align}
where $T_{1,\text{diel}}$ is the qubit energy relaxation time limitation by this dielectric loss.\\

In Supplementary Figure~\ref{fig:S21} \textbf{b}-\textbf{c}, the estimated $T_{1,\text{rad}}$ (dashed line) and $T_{1,\text{diel}}$ (dot-dashed line) as well as the resulting total energy relaxation time limitation $T_{1,\text{rad}}T_{1,\text{diel}}/(T_{1,\text{rad}}+T_{1,\text{diel}})$ (continuous line) are plotted vs. the area $A$ of the global gate which is connected \textbf{a} via a hypothetical $50\,\mathrm{\Omega}$ RF line ($Z=50\,\mathrm{\Omega}$), or \textbf{b} via a DC line as described above ($Z=400\,\mathrm{\Omega}$), or \textbf{c} in another hypothetical case when the gate is floating (we model the floating gate by choosing some huge number $Z\sim1\,\mathrm{G\Omega}$). Note in Supplementary Figure~\ref{fig:S21} that the dielectric loss in the filter capacitor (obviously) does not depend on the wiring impedance. The gate is circular and is placed centrally above the qubit. The distance between the qubit plane and the global gate is $d=0.9\,\mathrm{mm}$. In the experiment the gate diameter is $4\,\mathrm{mm}$, 
% see estimation C:\Arbeit\Google\sampleholder\Fotos\2019_strayJJ_Xmon_ChipXV_7\PC042346.svg
the corresponding gate area is indicated by a black dotted line. The qubit has an Xmon topology while its island area is $\sim2 \times w \times L$, where $w=30\,\mathrm{\mu m}$ is the qubit trace width (gap of $20\,\mathrm{\mu m}$), and $L=320\,\mathrm{\mu m}$ its length. The legend in Supplementary Figure~\ref{fig:S21} denotes various insulating film thicknesses. The cyan circle in \textbf{b} emphasizes the total $T_1$ limitation by the DC gate of about $6\,\mathrm{ms}$ in our experiment (insulating film thickness of $\sim60\,\mathrm{\mu m}$), which is dominated by dielectric loss.\\
\begin{figure*}[htbp]
	\begin{center}
		\includegraphics[width=.98\textwidth]{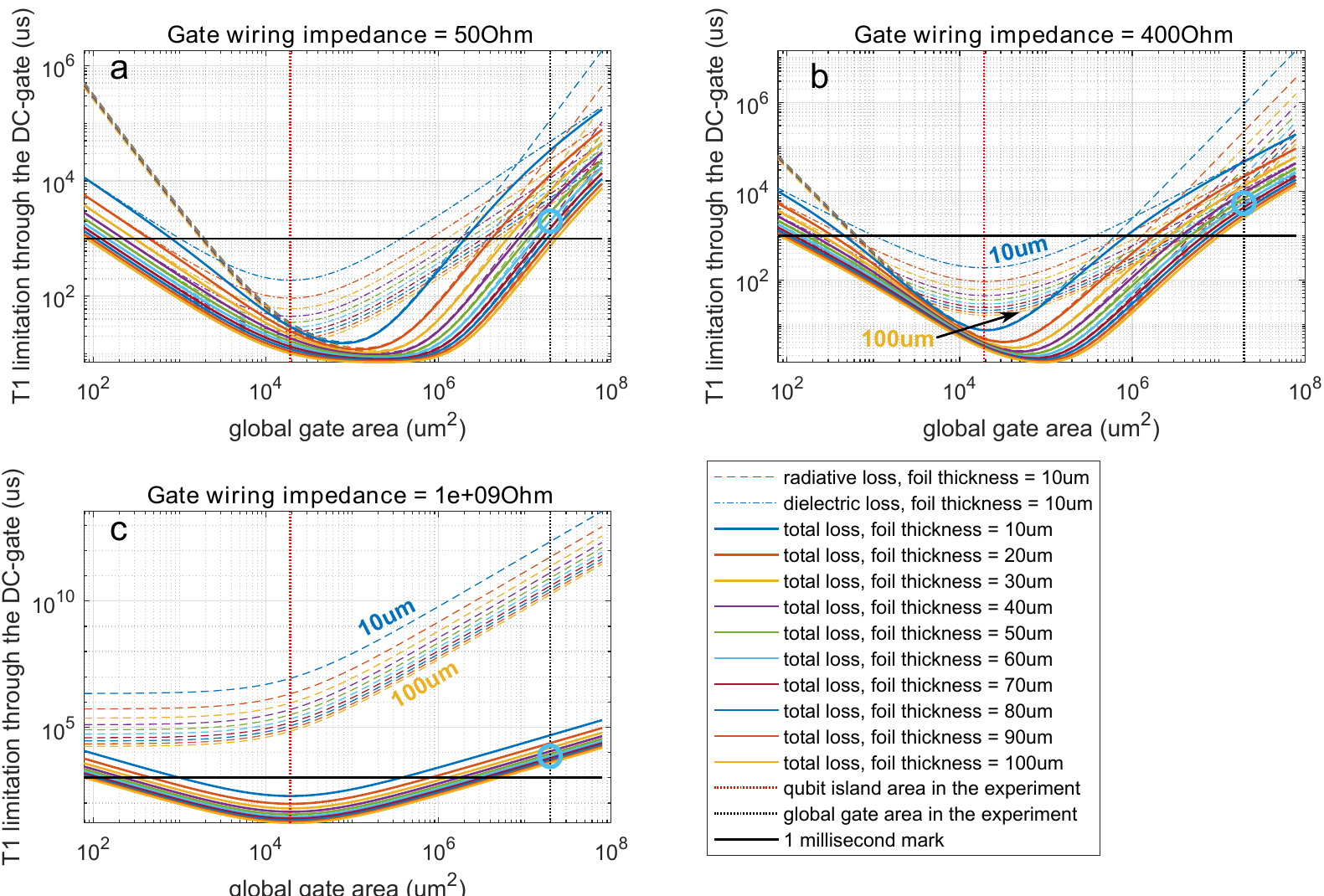} 
	\end{center}
	\caption{\textbf{$T_1$ limitation through the global DC gate.}
		Dielectric loss occurs in the insulating Kapton foil separating the DC-electrode from ground. Radiative loss depends on the impedance of the wiring connecting the gate electrode (see panel titles).
		\textbf{a} Hypothetical case if the DC gate was connected through an impedance-matched RF line. \textbf{b} $T_1$ estimation for the experiment reported here, when the DC gate is connected through a DC wire. \textbf{c} Hypothetical case if the DC gate was floating, which is modeled by a huge wire impedance of $1\,\mathrm{G\Omega}$.
	}
	\label{fig:S21}
\end{figure*}

\noindent \textbf{Notes}\\
The red dotted line in Supplementary Figure~\ref{fig:S21} \textbf{b} denotes the area of the qubit island. We see that the radiative losses are minimal if the global gate had a much smaller or a much larger footprint than the qubit island. For a small global gate, this is due to the small $C_\text{c}$ which maximizes $Q_\text{l}$ in ~\eqref{eq:Ql}. For large global gate areas, the filter capacitance is practically shorting the DC wire to ground. Similarly, the dielectric losses are rather weak for a small or a large global gate when the filter capacitor's participation ration $P_\text{f}$ becomes small. In Supplementary Figure~\ref{fig:S21} \textbf{a} a hypothetical case is presented if the global gate was connected through a $50\,\mathrm{\Omega}$-matched RF line. In such case one would expect increased radiative losses, however the total $T_1$ limitation would not go below $1\,\mathrm{ms}$. Another hypothetical scenario, which we consider to check the validitiy of our model, is presented in Supplementary Figure~\ref{fig:S21} \textbf{c}, where we consider the radiative loss if the global gate was floating (we model this by choosing a huge impedance for the gate wiring). As expected, we see that radiative losses then become negligible.\\

\section{Estimation of energy loss from a local DC gate}
Here, we estimate radiative dielectric loss from the individual DC-gate as proposed in the main text for integration into quantum processors (Fig. 4), and shown in Supplementary Figure~\ref{fig:S20}\textbf{b}. The wiring diagram is shown in Fig~\ref{fig:S20}\textbf{c}, it is identical to the setup with the global gate Supplementary Figure~\ref{fig:S20}\textbf{a}. The only difference is that in the flip-chip architecture, the Xmon's arm length needs to be shortened to $\sim300\,\mathrm{\mu m}$ in order to keep the qubit total shunt capacitance at $\sim97\,\mathrm{fF}$. The energy relaxation time limitation due to radiative loss is given in Eq.~\eqref{eq:Ql}, and that due to dielectric loss in the AlO$_\text{x}$ spacer (assuming $\tan\delta_\mathrm{AlO_x}\sim10^{-3}$) is given in Eq.~\eqref{eq:t1diel}. 

The $T_1$ limitation is shown as a function of the local gate in Supplementary Figure~\ref{fig:S22}. In \textbf{a}, \textbf{b} and \textbf{c}, we consider the cases of the local gate connected via a $50\,\mathrm{\Omega}$-matched RF line, a high-impedance DC line, or a 1 G$\Omega$ impedance, respectively. The orange circle in \textbf{b} denotes the total $T_1$ limitation of about $3\,\mathrm{ms}$ when the DC gate is controlled via a DC-wire, and the AlO$_\text{x}$ has a realistic thickness of $25\,\mathrm{nm}$. Shown in \textbf{c} is the case when the local gate is assumed to be floating, which results in negligible radiative loss.\\

\begin{figure*}[htbp]
	\begin{center}
		\includegraphics[width=.98\textwidth]{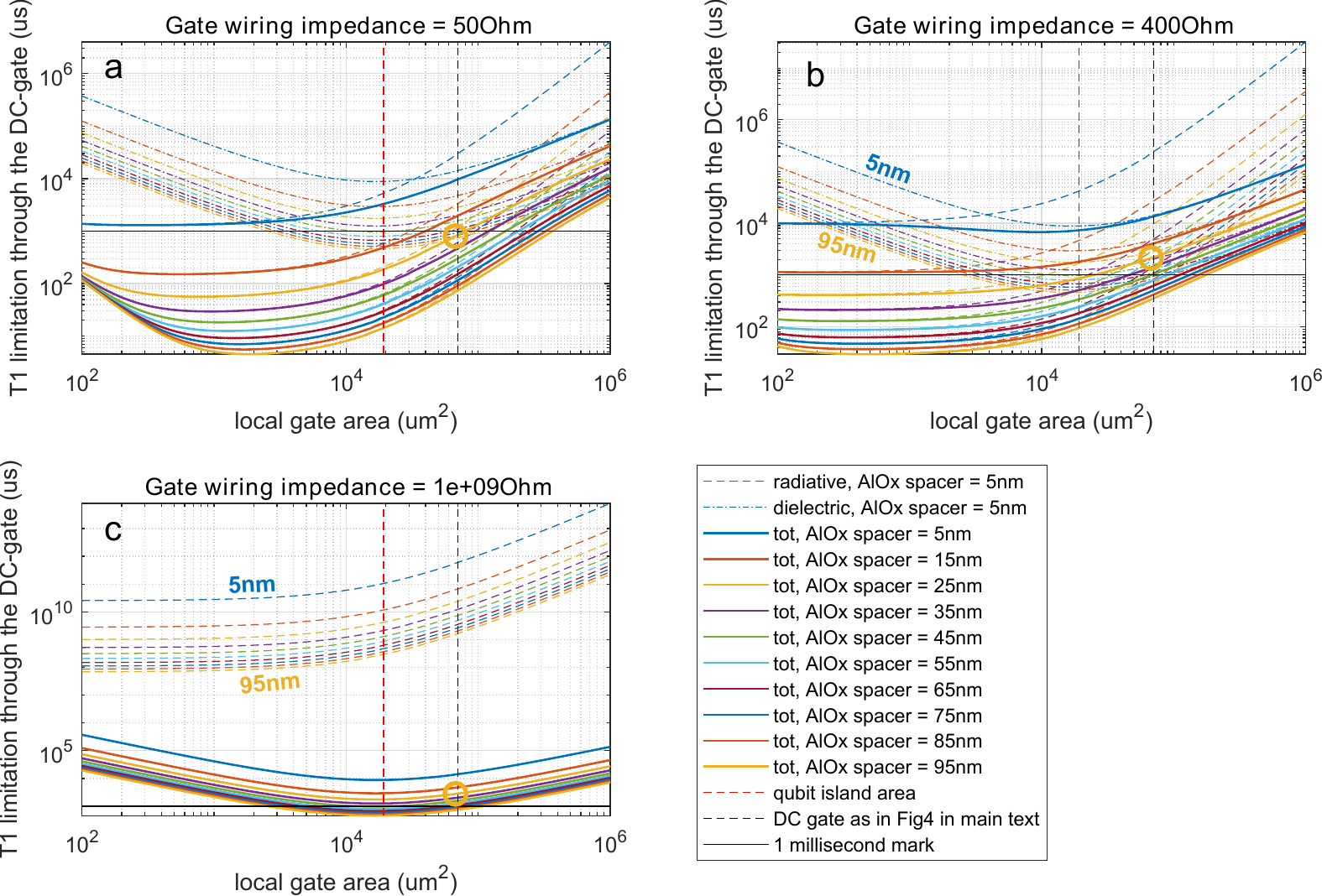} 
	\end{center}
	\caption{\textbf{$T_1$ limitation through a local DC gate as proposed in the main text (Fig 4).}
		\textbf{a} Hypothetical case if the DC gate was connected through an impedance-matched RF line. \textbf{b} $T_1$ estimation for the proposed setup, when the DC gate is connected through a DC wire. \textbf{c} Hypothetical case if the DC gate was floating, which is modeled by a huge wire impedance of $1\,\mathrm{G\Omega}$.
	}
	\label{fig:S22}
\end{figure*}

\section{$T_1$-enhancement as a function of time}
\begin{figure*}[htbp]
	\begin{center}
		\includegraphics[width=.98\textwidth]{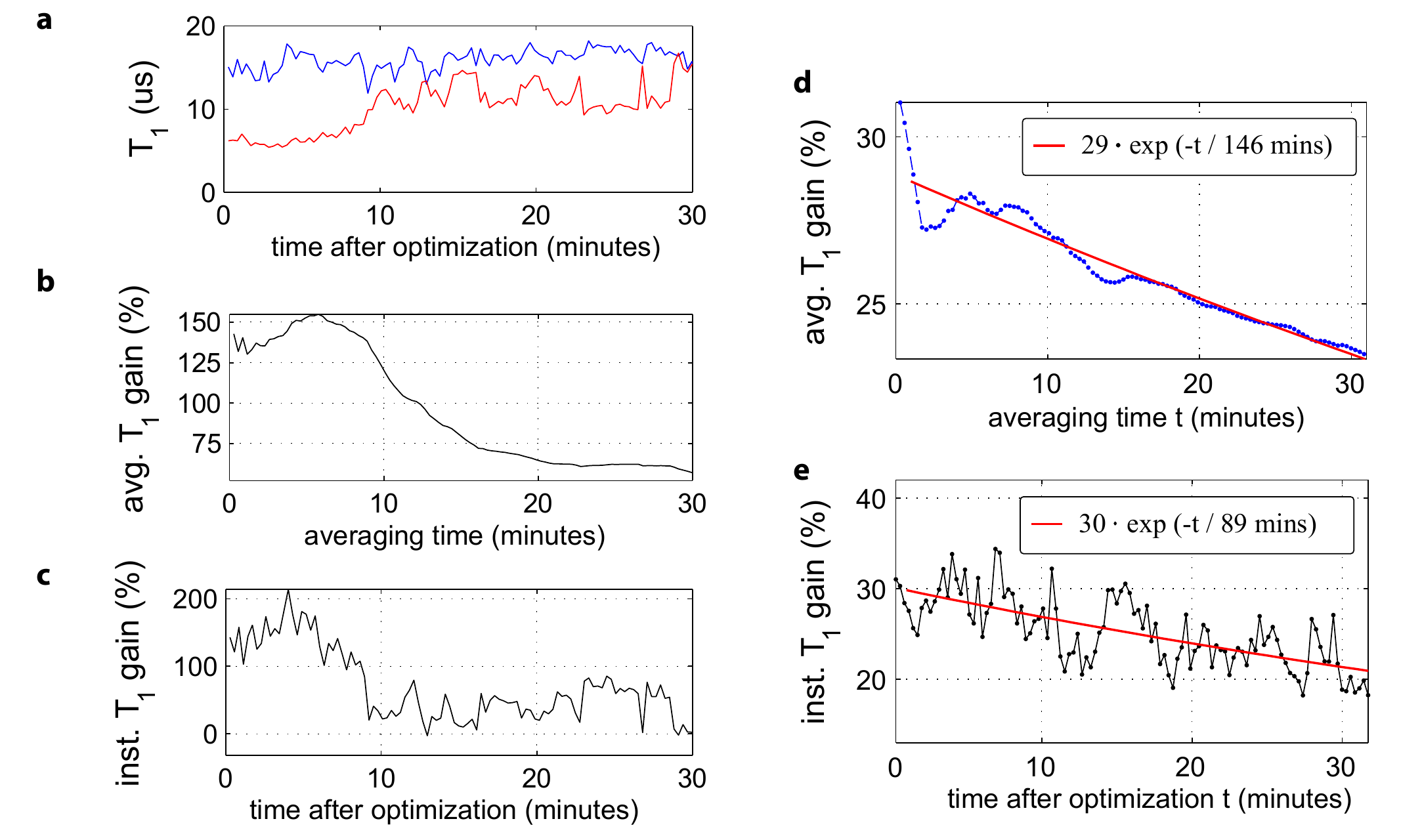} 
	\end{center}
	\caption{\textbf{$T_1$ gain as a function of time.}
		\textbf{a}  Example measurement of qubit $T_1$ times before (red) and after (blue) optimization, each observed during 30 minutes.
		\textbf{b} Corresponding average $T_1$ gain, averaging instantaneous $T_1$ gains since the time of optimization.
		\textbf{c} Instantaneous $T_1$ gain at a given time after optimization. Note that it may drop to zero in this example, in contrast to the average $T_1$ gain shown in \textbf{b}.  
		\textbf{d} average $T_1$ gain and \textbf{e} instantaneous $T_1$ gain, both averaged over all repetitions taken at different qubit frequencies. The red lines are exponential fits as indicated in the legends.
	}
	\label{fig:T1gain}
\end{figure*}

As a consequence of the defect's resonance frequency fluctuations, the enhancement (gain) of the $T_1$ time tends to diminish with time that has passed after the E-field optimization. Here, we characterize two types of $T_1$ gain: the \emph{instantaneous $T_1$ gain} is the individual $T_1$ improvement measured at the corresponding time $t$:
\[
\mathrm{inst. T_1\ gain}(t) = [T_\mathrm{1,opt}(t) -T_\mathrm{1,reg}(t)] / T_\mathrm{1,reg}(t),
\]
where $T_\mathrm{1,reg}(t)$ is the $T_1$ time measured at zero electric field at the time $t$ after the beginning of the 30-minute observation interval, and $T_\mathrm{1,opt}(t)$ is the $T_1$ time measured at optimized electric field at the time $t$ after the optimization was done. The \emph{average $T_1$ gain} denotes the average over the instantaneous $T_1$ improvements up to a certain time since optimization.
To illustrate this, Supplementary Figure~\ref{fig:T1gain} shows exemplary data of $T_1$ times measured before and after optimization, and the corresponding average and instantaneous $T_1$ gains.\\

Figures~\ref{fig:T1gain}\textbf{d} and \textbf{e} show average and instantaneous $T_1$ gains, respectively, that are averaged over all repetitions taken at 60 different qubit frequencies. Both average and instantaneous $T_1$ gain drop from initial values of ~30\% directly after optimization to slightly above 20\% after 30 minutes past optimization. To get a feeling for the time scale on which the $T_1$ gain dwindles, we added fits to an exponential decay law: $T_1\ \mathrm{gain}(t) = A \cdot \mathrm{exp} (-t / B)$ (red lines in Figs.~\ref{fig:T1gain}\textbf{d} and \textbf{e}), which result in $A$=29 (30) and $B$=146 (89) minutes for average (instantaneous) $T_1$ gains.

\clearpage
\section{Additional data}
\begin{figure*}[htb!]
	%	\begin{center}
		\includegraphics[width=\textwidth, height=11cm]{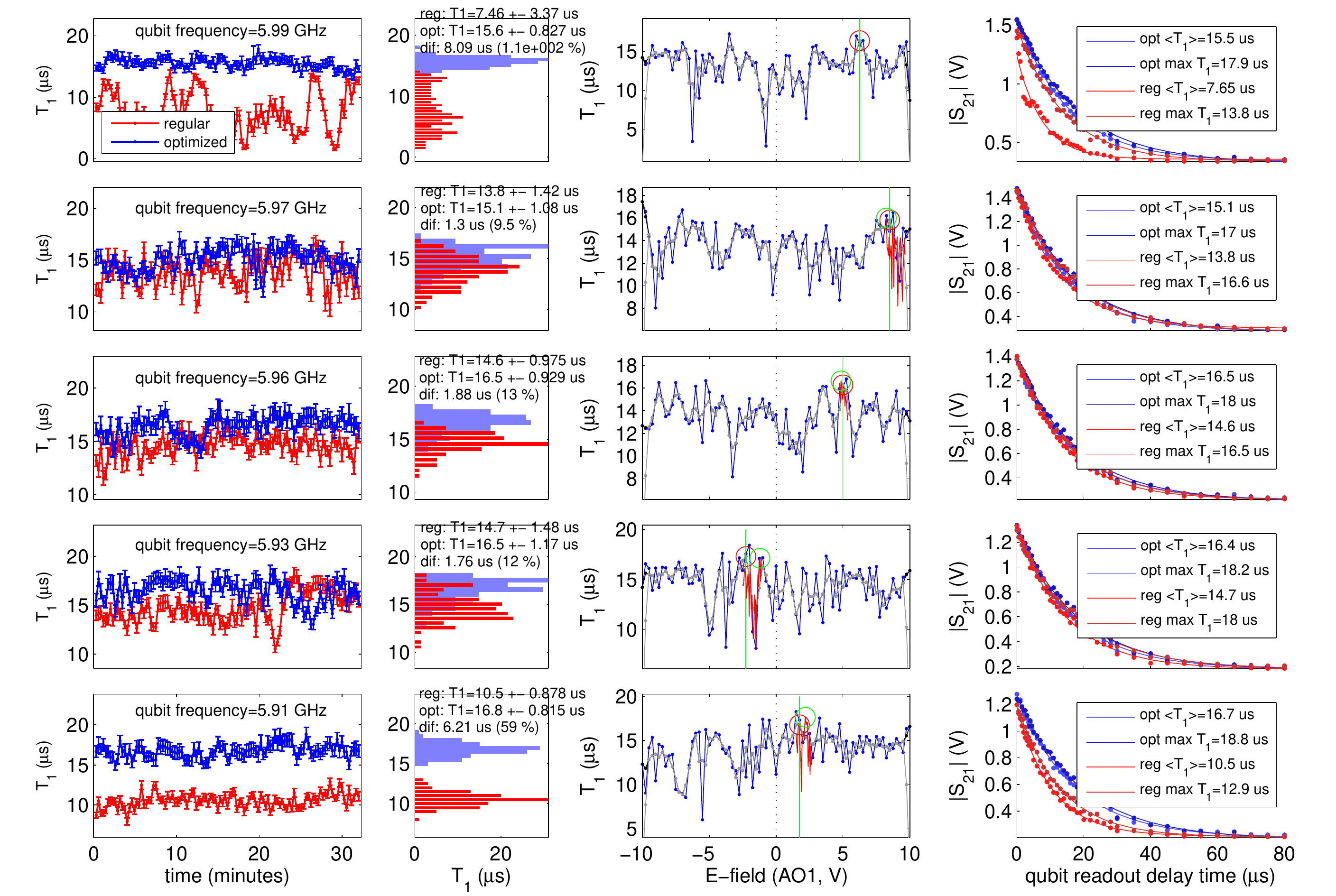}
		\includegraphics[width=\textwidth, height=11cm]{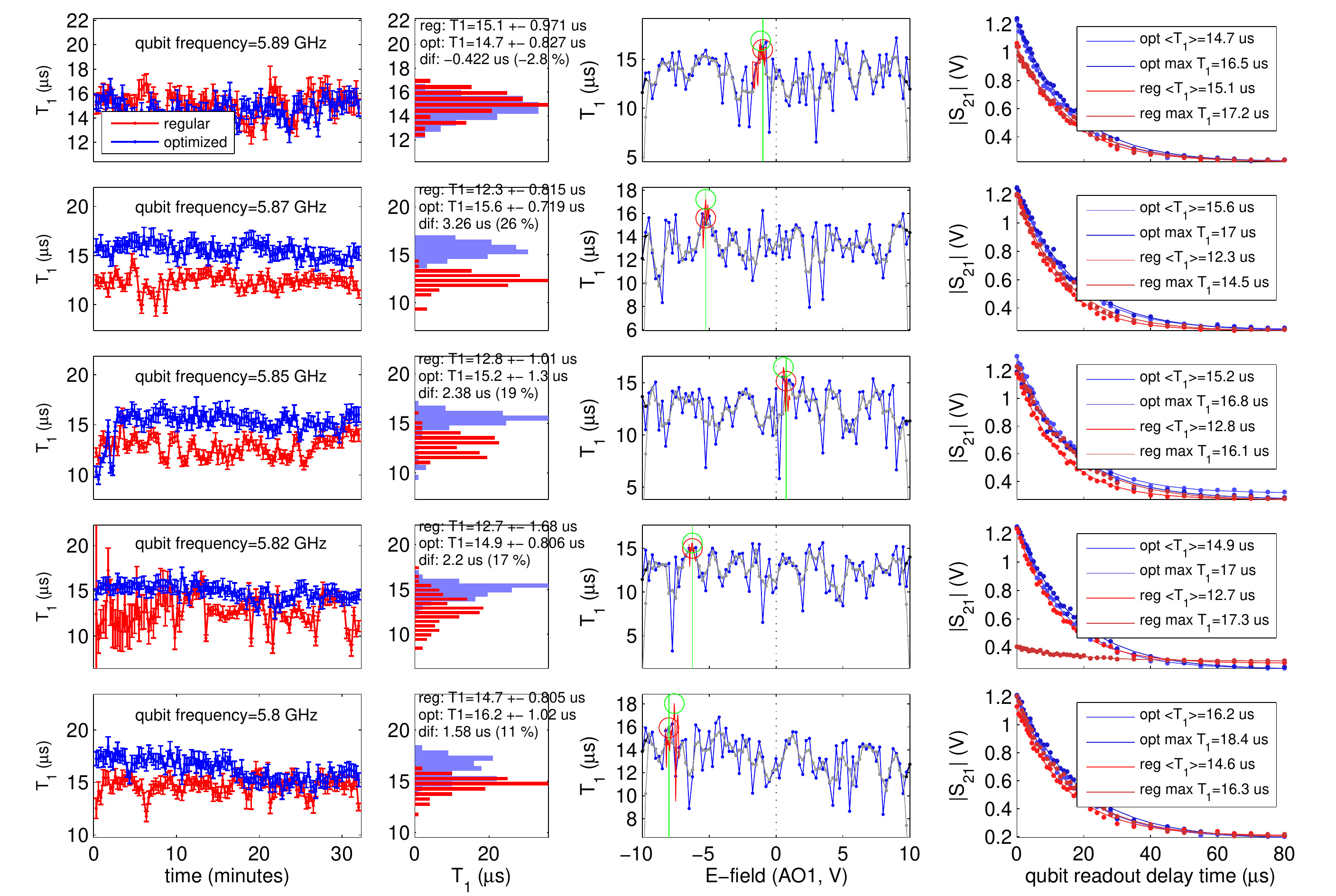}
		%	\end{center}
	\caption{Testing the optimization routine at various qubit frequencies (rows).
		\textbf{Column 1:} $T_1$ time measured for 30 minutes at zero E-field (red) and optimized E-field (blue). \textbf{Column 2:} Histograms of $T_1$ during 30 minutes for optimized (blue) and zero E-field (red). \textbf{Column 3:} $T_1$ vs. applied E-field to find the optimum E-field (red circle). Data obtained in the second pass is shown in red. \textbf{Column 4:} Examples of raw qubit decay curves showing a mean ($<T_1>$) and maximum (max)  $T_1$ time acquired at optimized (blue) and zero (red) applied E-field.		
	}
	\label{fig:sm1}
\end{figure*}

\begin{figure*}[htb!]
	%	\begin{center}
		\includegraphics[width=\textwidth, height=11cm]{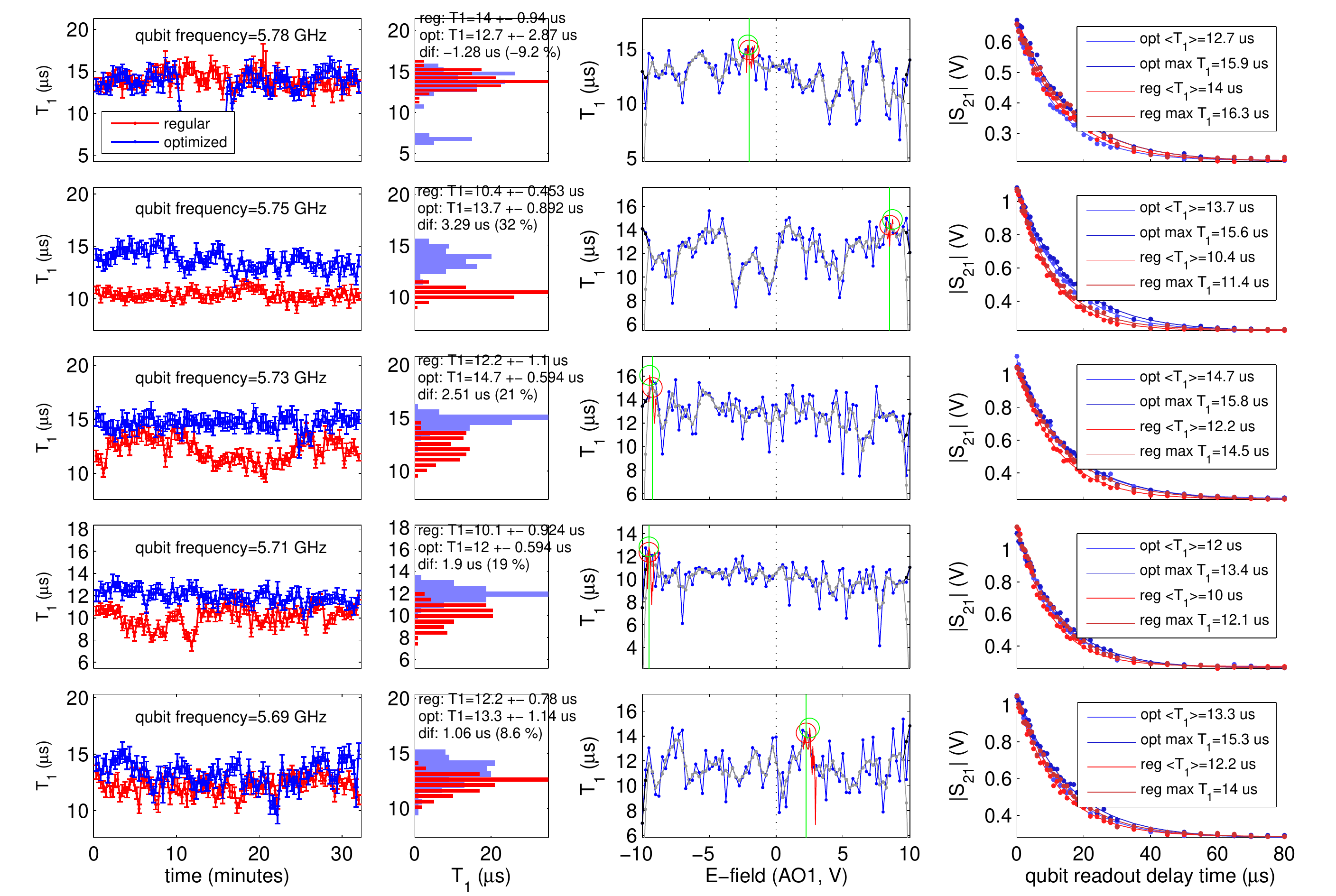}
		\includegraphics[width=\textwidth, height=11cm]{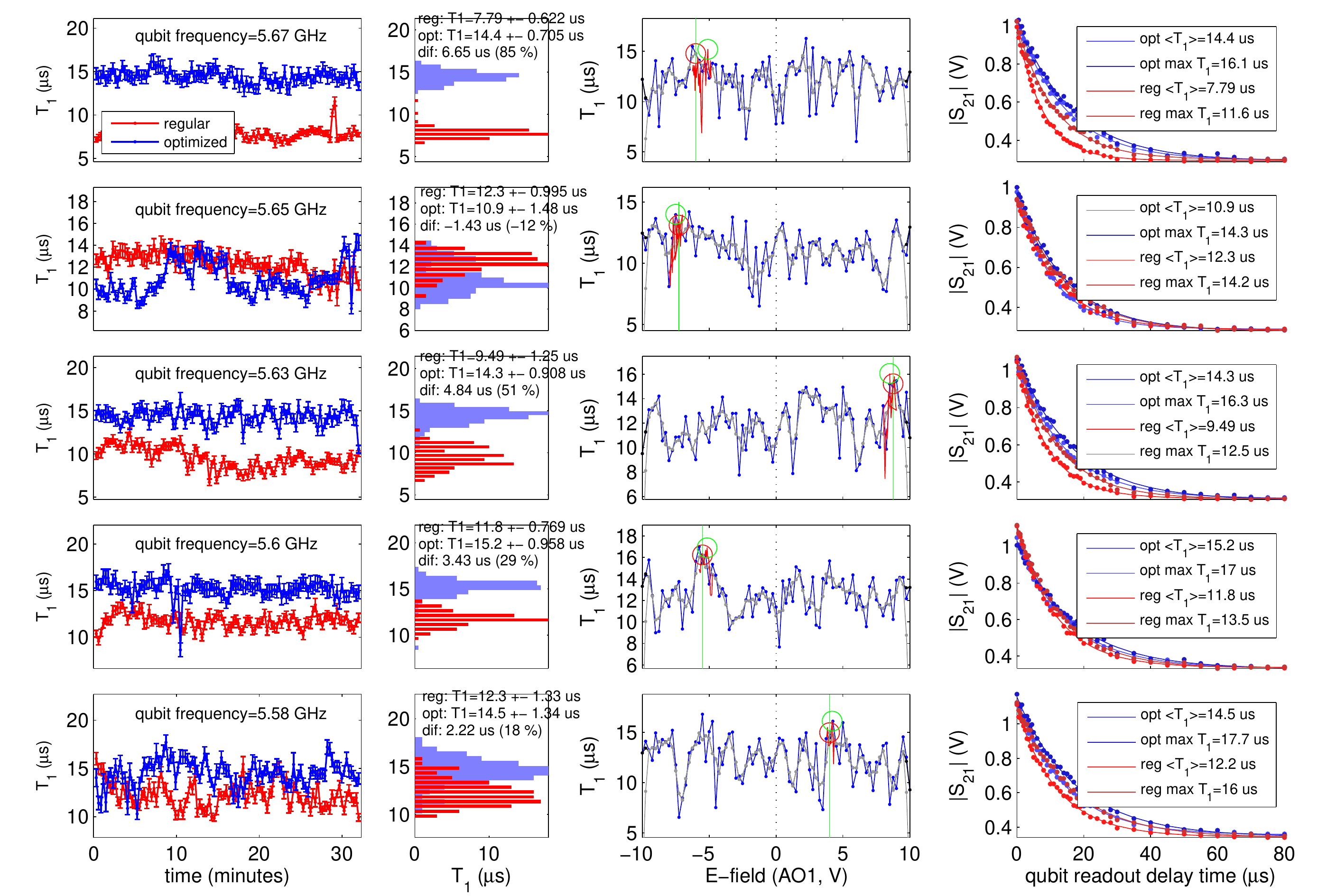}
		%	\end{center}
	\caption{Testing the optimization routine at various qubit frequencies (rows).
		\textbf{Column 1:} $T_1$ time measured for 30 minutes at zero E-field (red) and optimized E-field (blue). \textbf{Column 2:} Histograms of $T_1$ during 30 minutes for optimized (blue) and zero E-field (red). \textbf{Column 3:} $T_1$ vs. applied E-field to find the optimum E-field (red circle). Data obtained in the second pass is shown in red. \textbf{Column 4:} Examples of raw qubit decay curves showing a mean ($<T_1>$) and maximum (max)  $T_1$ time acquired at optimized (blue) and zero (red) applied E-field.		
	}
	\label{fig:sm2}
\end{figure*}

\begin{figure*}[htb!]
	%	\begin{center}
		\includegraphics[width=\textwidth, height=11cm]{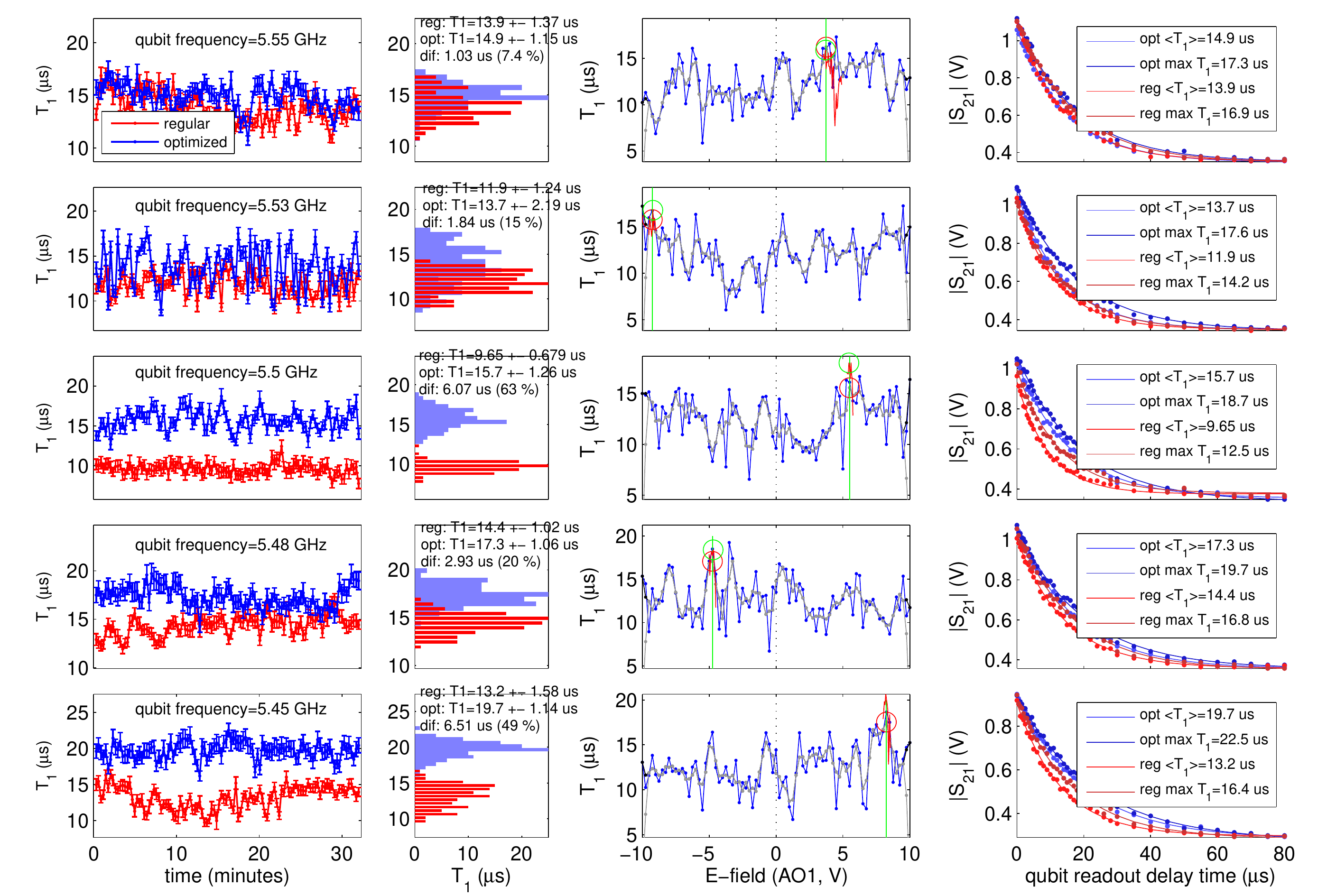}
		\includegraphics[width=\textwidth, height=11cm]{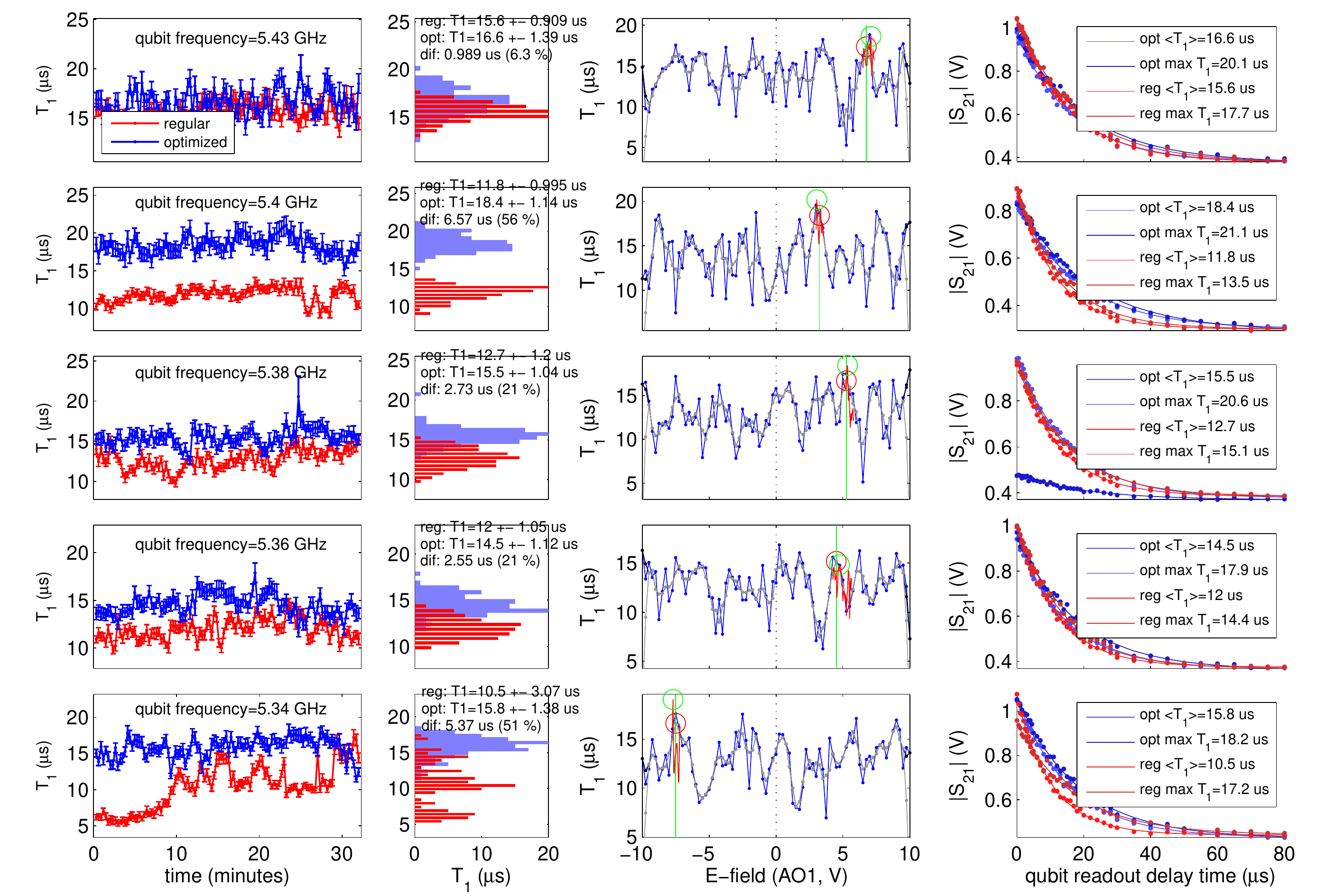}
		%	\end{center}
	\caption{Testing the optimization routine at various qubit frequencies (rows).
		\textbf{Column 1:} $T_1$ time measured for 30 minutes at zero E-field (red) and optimized E-field (blue). \textbf{Column 2:} Histograms of $T_1$ during 30 minutes for optimized (blue) and zero E-field (red). \textbf{Column 3:} $T_1$ vs. applied E-field to find the optimum E-field (red circle). Data obtained in the second pass is shown in red. \textbf{Column 4:} Examples of raw qubit decay curves showing a mean ($<T_1>$) and maximum (max)  $T_1$ time acquired at optimized (blue) and zero (red) applied E-field.		
	}
	\label{fig:sm3}
\end{figure*}

\begin{figure*}[htb!]
	%	\begin{center}
		\includegraphics[width=\textwidth, height=11cm]{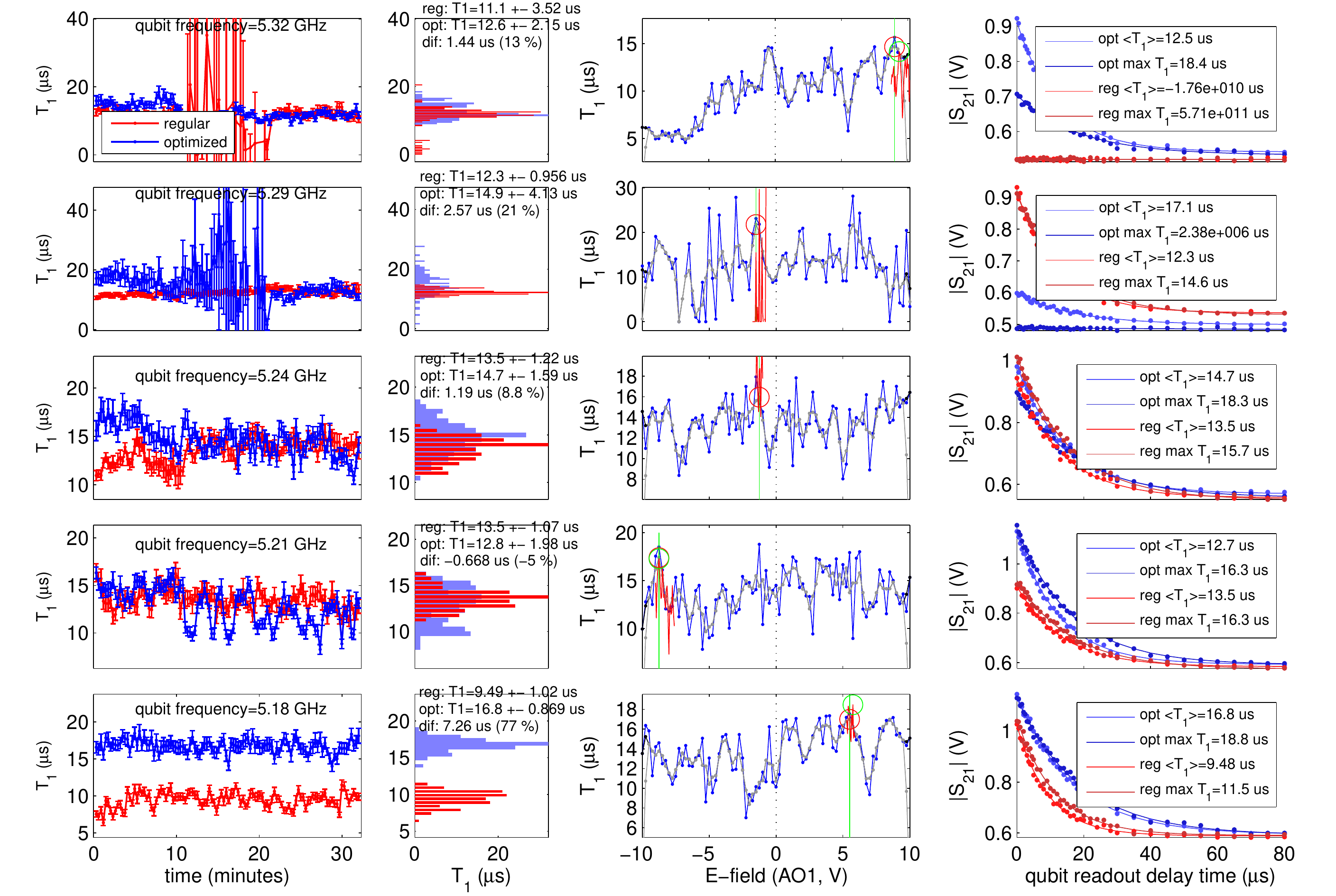}
		\includegraphics[width=\textwidth, height=11cm]{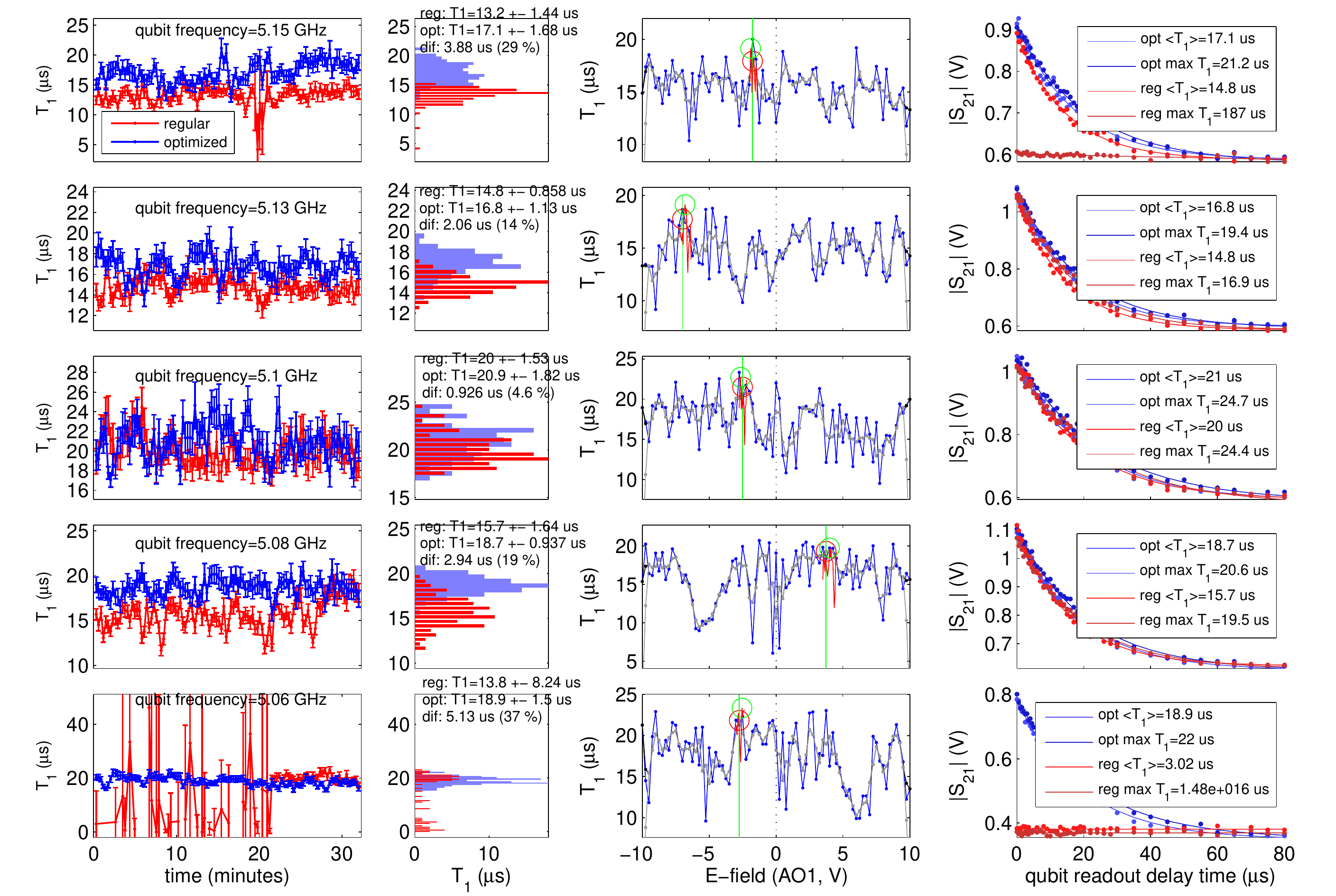}
		%	\end{center}
	\caption{Testing the optimization routine at various qubit frequencies (rows).
		\textbf{Column 1:} $T_1$ time measured for 30 minutes at zero E-field (red) and optimized E-field (blue). \textbf{Column 2:} Histograms of $T_1$ during 30 minutes for optimized (blue) and zero E-field (red). \textbf{Column 3:} $T_1$ vs. applied E-field to find the optimum E-field (red circle). Data obtained in the second pass is shown in red. \textbf{Column 4:} Examples of raw qubit decay curves showing a mean ($<T_1>$) and maximum (max)  $T_1$ time acquired at optimized (blue) and zero (red) applied E-field.		
	}
	\label{fig:sm4}
\end{figure*}

\begin{figure*}[htb!]
	%	\begin{center}
		\includegraphics[width=\textwidth, height=11cm]{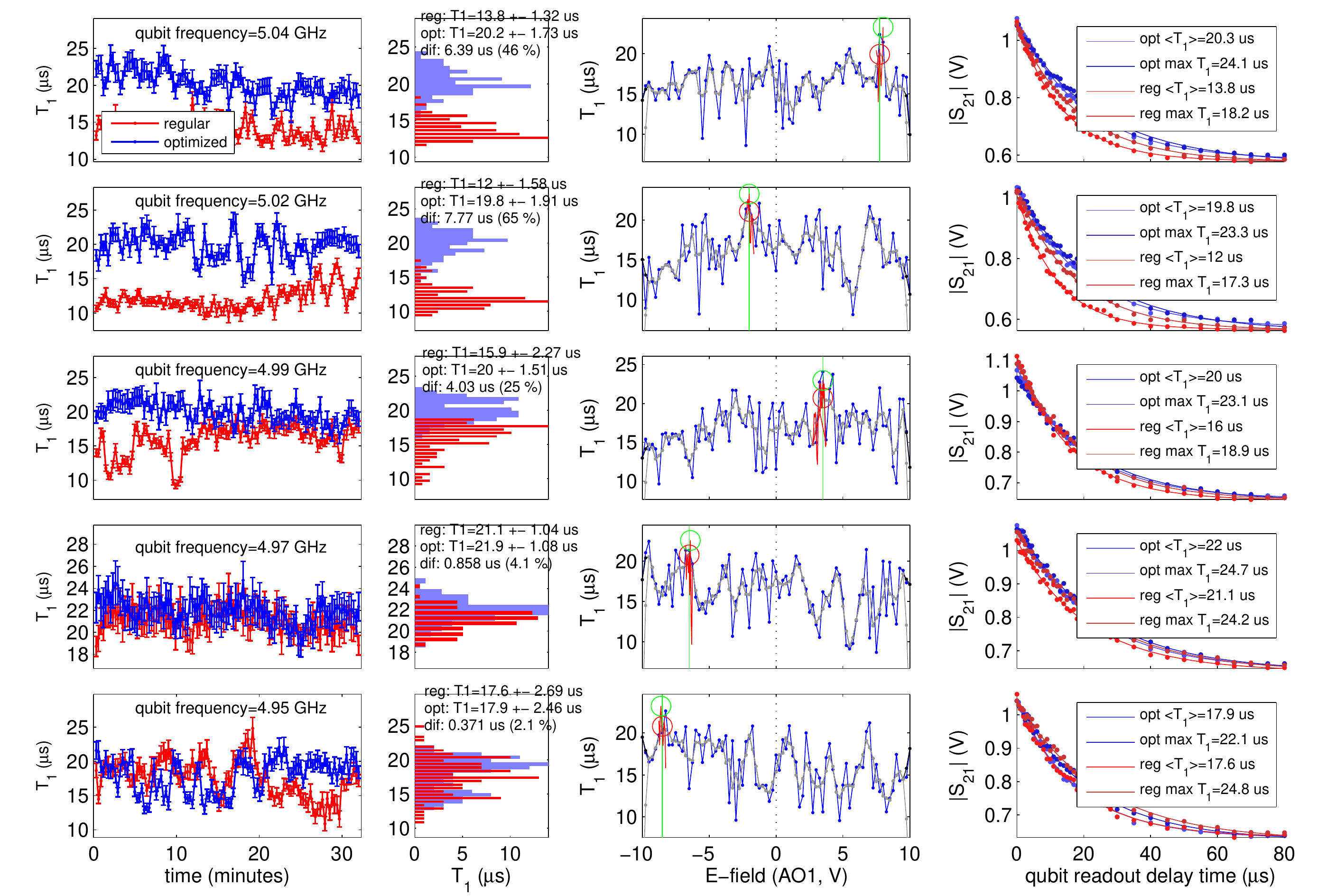}
		\includegraphics[width=\textwidth, height=11cm]{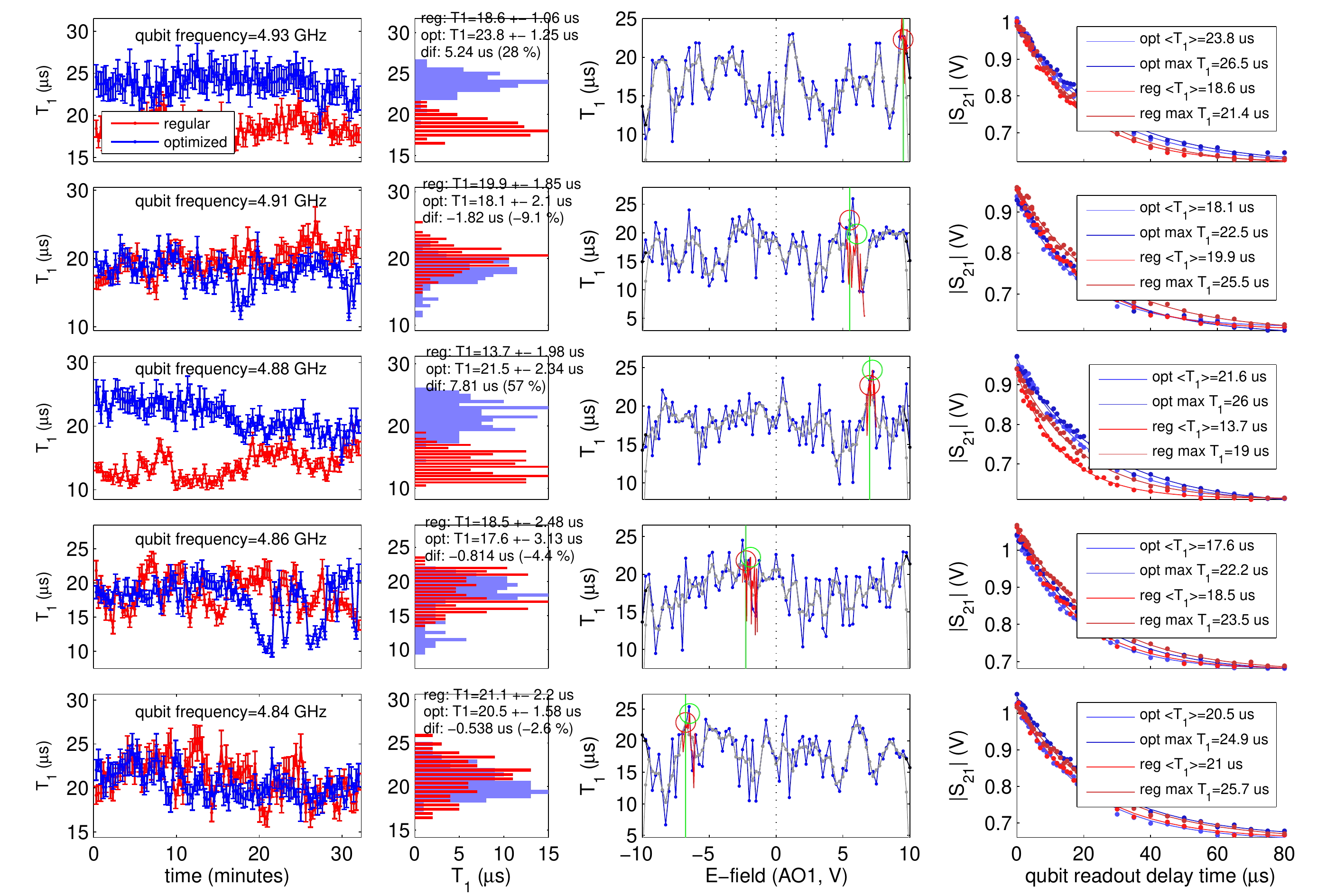}
		%	\end{center}
	\caption{Testing the optimization routine at various qubit frequencies (rows).
		\textbf{Column 1:} $T_1$ time measured for 30 minutes at zero E-field (red) and optimized E-field (blue). \textbf{Column 2:} Histograms of $T_1$ during 30 minutes for optimized (blue) and zero E-field (red). \textbf{Column 3:} $T_1$ vs. applied E-field to find the optimum E-field (red circle). Data obtained in the second pass is shown in red. \textbf{Column 4:} Examples of raw qubit decay curves showing a mean ($<T_1>$) and maximum (max)  $T_1$ time acquired at optimized (blue) and zero (red) applied E-field.		
	}
	\label{fig:sm5}
\end{figure*}

\begin{figure*}[htb!]
	%	\begin{center}
		\includegraphics[width=\textwidth, height=11cm]{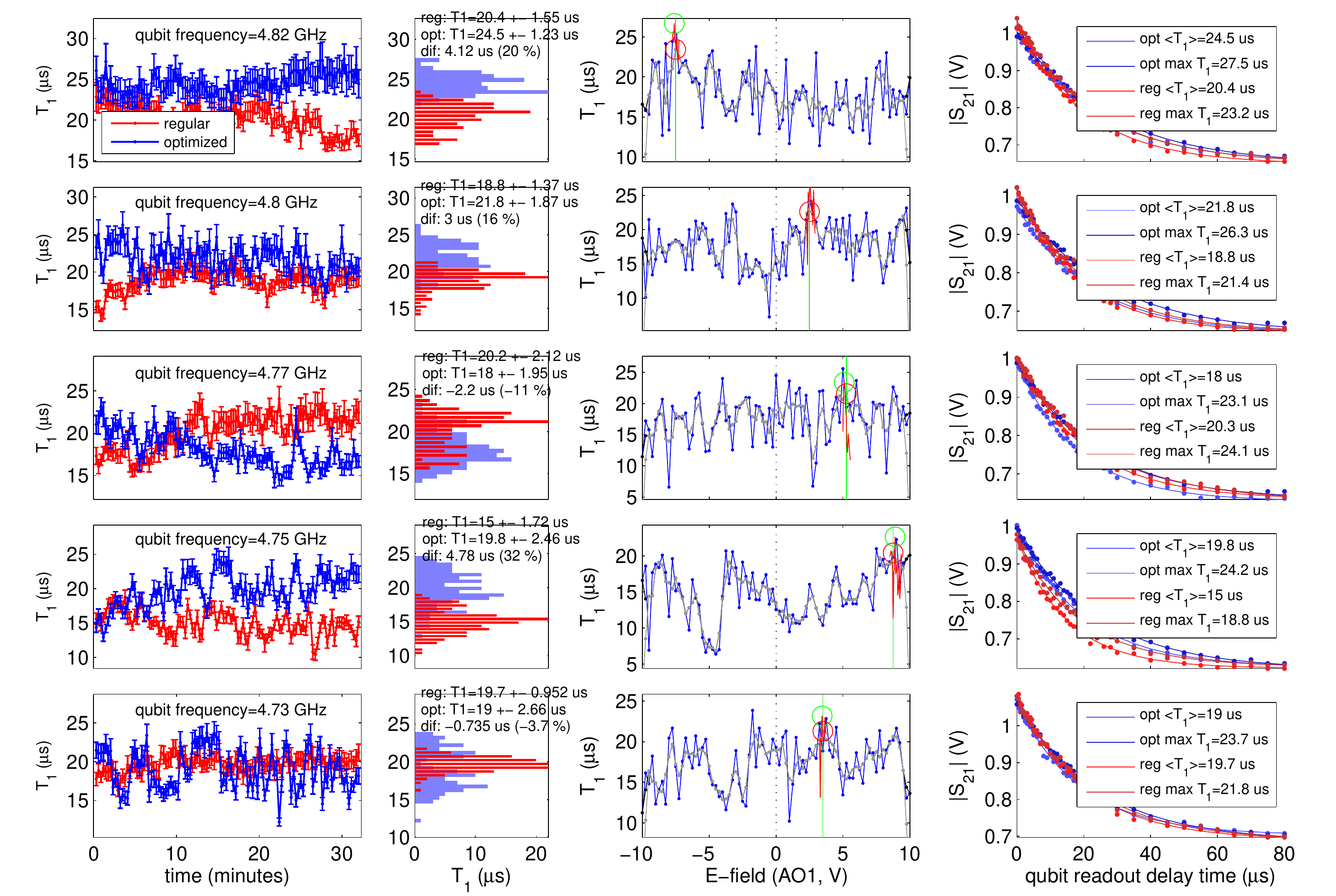}
		\includegraphics[width=\textwidth, height=11cm]{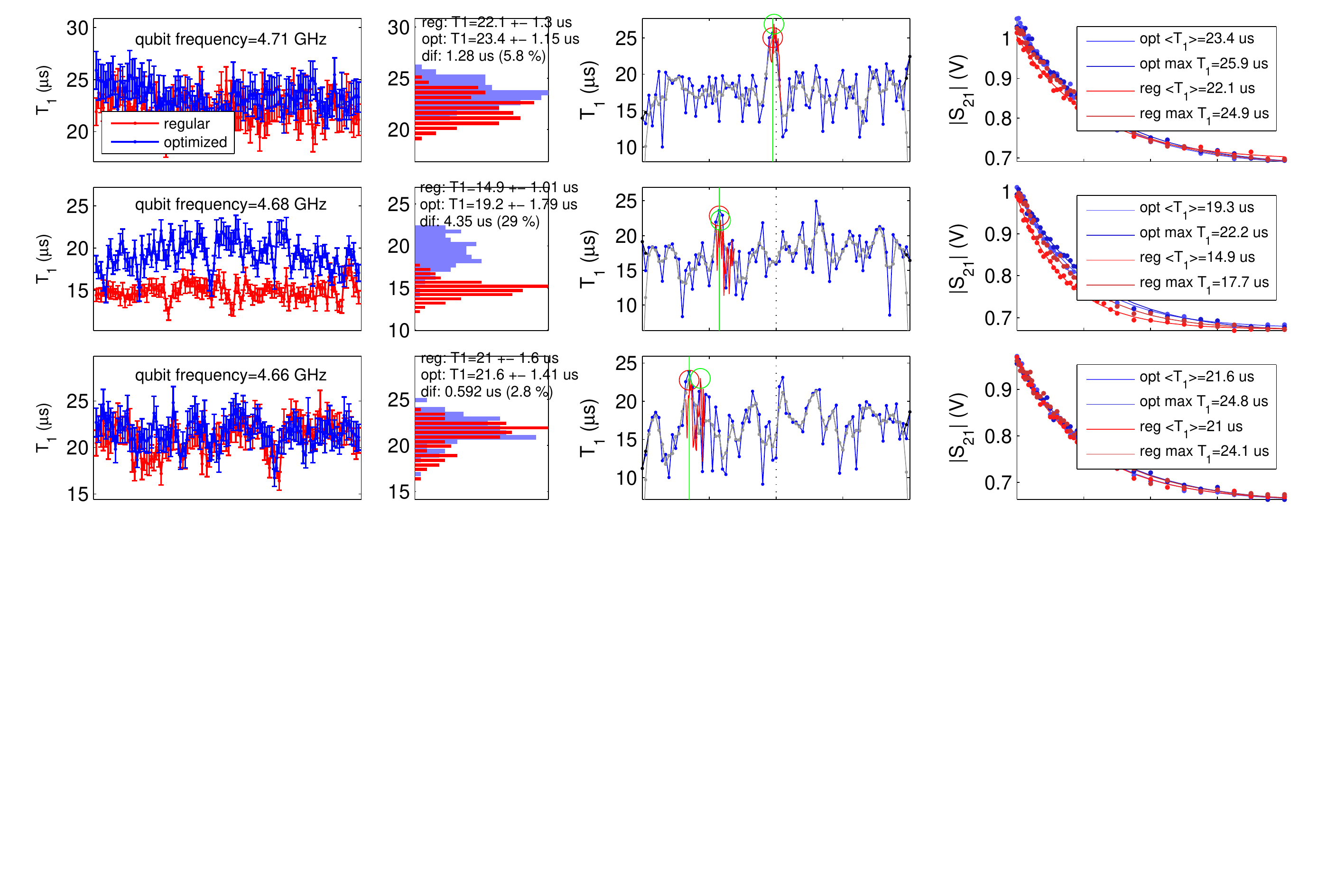}
		%	\end{center}
	\caption{Testing the optimization routine at various qubit frequencies (rows).
		\textbf{Column 1:} $T_1$ time measured for 30 minutes at zero E-field (red) and optimized E-field (blue). \textbf{Column 2:} Histograms of $T_1$ during 30 minutes for optimized (blue) and zero E-field (red). \textbf{Column 3:} $T_1$ vs. applied E-field to find the optimum E-field (red circle). Data obtained in the second pass is shown in red. \textbf{Column 4:} Examples of raw qubit decay curves showing a mean ($<T_1>$) and maximum (max)  $T_1$ time acquired at optimized (blue) and zero (red) applied E-field.		
	}
	\label{fig:sm6}
\end{figure*}

Supplementary Figures~\ref{fig:sm1} - \ref{fig:sm6} show the full data set acquired from the optimization routine at various qubit resonance frequencies.\\

\end{document}